\documentclass[aps,prc,twocolumn,showpacs,showkeys,superscriptaddress]{revtex4}
\usepackage{longtable}
\usepackage{graphicx}
\usepackage{dcolumn}
\usepackage{color}
\usepackage{float}
\begin{document}
\title{Identification of a long lived $\beta$ decaying isomer in $^{150}$Pm}
\author{A.~Saha}
\affiliation{Variable Energy Cyclotron Centre, Kolkata - 700 064, India}
\affiliation{Homi Bhabha National Institute, Training School Complex,
             Anushakti Nagar, Mumbai - 400\,094, India}
\author{T.~Bhattacharjee}
\thanks{Corresponding author}
\email{btumpa@vecc.gov.in}
\affiliation{Variable Energy Cyclotron Centre, Kolkata - 700 064, India}
\affiliation{Homi Bhabha National Institute, Training School Complex,
             Anushakti Nagar, Mumbai - 400\,094, India}
\author{D.~Banerjee}
\affiliation{RCD-BARC, Variable Energy Cyclotron Centre, Kolkata - 700 064, India}
\author{S.~S.~Alam}
\affiliation{Variable Energy Cyclotron Centre, Kolkata - 700 064, India}
\affiliation{Homi Bhabha National Institute, Training School Complex,
             Anushakti Nagar, Mumbai - 400\,094, India}
\author{ Deepak~Pandit}
\affiliation{Variable Energy Cyclotron Centre, Kolkata - 700 064, India}
\author{M.~Saha~Sarkar}
\affiliation{Saha Institute of Nuclear Physics, Kolkata - 700 064, India}
\author{S.~Sarkar}
\affiliation{Indian Institute of Engineering Science and Technology, Shibpur, West Bengal - 711 103, India}
\author{ P.~Das}
\affiliation{Variable Energy Cyclotron Centre, Kolkata - 700 064, India}
\affiliation{Homi Bhabha National Institute, Training School Complex,
             Anushakti Nagar, Mumbai - 400\,094, India}
\author{Soumik Bhattacharya}
\affiliation{Variable Energy Cyclotron Centre, Kolkata - 700 064, India}
\affiliation{Homi Bhabha National Institute, Training School Complex,
             Anushakti Nagar, Mumbai - 400\,094, India}
\author{R.~Guin}
\affiliation{RCD-BARC, Variable Energy Cyclotron Centre, Kolkata - 700 064, India}
\author{S.~K.~Das}
\affiliation{RCD-BARC, Variable Energy Cyclotron Centre, Kolkata - 700 064, India}
\author{ S.~R.~Banerjee}
\affiliation{Variable Energy Cyclotron Centre, Kolkata - 700 064, India}
\affiliation{Homi Bhabha National Institute, Training School Complex,
             Anushakti Nagar, Mumbai - 400\,094, India}
\date{\today}

\begin{abstract}

The decay of odd-odd $^{150}$Pm has been studied by populating the nucleus with the $^{150}$Nd(p,n)$^{150}$Pm reaction at E$_{beam}$ = 8.0 MeV using 97$\%$ enriched $^{150}$Nd target. The presence of an isomeric state with $\beta$ decay half life of 2.2(1) h  could be identified in $^{150}$Pm by following the half lives of the observed $\gamma$ transitions. The decay of the isomer to the excited levels of $^{150}$Sm has been confirmed by observing the $\gamma - \gamma$ coincidence with the VENUS array of six Compton suppressed Clover HPGe detectors. The $\beta$ decay end-point energies corresponding to the decay from the $^{150g}$Pm and $^{150m}$Pm have been measured using a $\beta-\gamma$ coincidence setup of two thin window Planar HPGe detectors and four Clover HPGe detectors of the VENUS array. The systematics of the similar isomeric states in neighboring nuclei has been studied to understand the underlying structure of these states. Shell model calculation has been performed by using OXBASH code which indicates the presence of a 5$^-$ isomeric state at very low excitation in the nucleus. The calculation also suggests hindered electromagnetic decay of this isomer and supports the possibility of its $\beta$ decay to the excited levels of $^{150}$Sm.

\end{abstract}

\pacs{23.20.Lv; 23.40.−s; 21.10.Tg; 21.60.Cs}

\keywords{Nuclear Structure; Decay $\gamma$ spectroscopy; $\beta$ decay; HPGe detector; Shell Model Calculation.}

\maketitle

\section{Introduction}
\label{intro}

The odd-odd $^{150}$Pm nucleus lies in the transitional region between N = 88 and N = 90 with three proton holes compared to the subshell closure of Z = 64. The ground state shape transition as a function of neutron number, associated with the softness of the collective potential, has been observed in this region and remained a topic of both experimental and theoretical interest~\cite{iachello,casten,meyer,garret,nomura,humby}. Similar shape transition is also indicated in case of odd-A Pm nuclei from the results obtained using neutron proton Interacting Boson Fermion Model calculation~(IBFM)~\cite{ibm-pm}. In this work, it has been observed that the experimental energy spectra of $^{147}$Pm and $^{149}$Pm nuclei could be reproduced by coupling the odd proton with a vibrational core whereas a rotational core was required for $^{151}$Pm and $^{153}$Pm.

\begin{figure*}
  \begin{center}
   \includegraphics[width=\textwidth, angle=0]{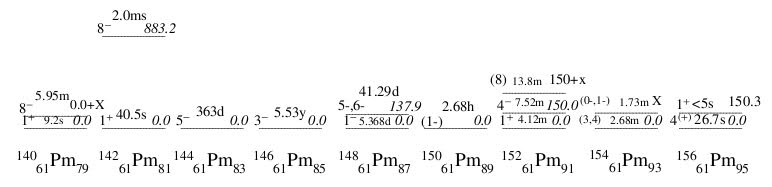}
  \caption{The systematics of the isomeric states in Z = 61 isotopes neighboring to $^{150}$Pm. The data is taken from ENSDF database~\cite{ensdf}. The J$^{\pi}$ of the ground state in $^{150}$Pm is suggested to be 2$^-$ of ~\cite{pm150-prc}.}
  \label{fig0a-syst}
  \end{center}
  \end{figure*}

\begin{figure*}
  \begin{center}
   \includegraphics[width=\textwidth, angle=0]{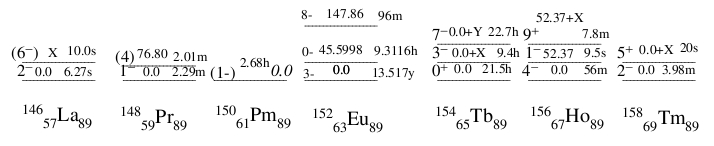}
  \caption{The systematics of the isomeric states in N = 89 isotones neighboring to $^{150}$Pm. The data is taken from ENSDF database~\cite{ensdf}. The J$^{\pi}$ of the ground state of $^{150}$Pm is suggested to be 2$^-$ in ~\cite{pm150-prc}.}
  \label{fig0b-syst}
  \end{center}
  \end{figure*}

In the odd-odd nuclei of this mass region, several long lived isomeric states are known and predicted~\cite{jain1}. The systematics of the low lying states in the odd-odd nuclei around $^{150}$Pm also reveal that the said isomers are observed in many of these nuclei~\cite{ensdf-a146,ensdf-a148,ensdf-a152,ensdf-a154,ensdf-a156}, as shown in Fig.~\ref{fig0a-syst} and~\ref{fig0b-syst}. However, no such isomeric state is known till date in case of $N = 89$ Pm nucleus~\cite{ensdf-a150}. These isomers observed in the deformed odd-odd nuclei around $^{150}$Pm are mostly understood on the basis of Gallagher Moszkowski (GM) rule~\cite{gmrule} for coupling of angular momentum based on the Nilsson configurations of the neighboring odd Z and odd N nuclei. However, this type of calculation is not appropriate for the nuclei having very low deformation where the shell model calculations might be the right choice. Also, the existing knowledge about the structure and the associated configurations for the low lying levels in the neighboring nuclei is very scanty~\cite{ensdf-a147,ensdf-a149,ensdf-a151,ensdf-a153}. Hence, the systematic understanding of these isomers is challenging in the context of choosing the appropriate model based calculation and also the available information in the neighboring nuclei. This might be the most probable reason for the fact that the nucleonic configurations of the two known isomers in $^{152}$Pm are unassigned till date.

Most of the long lived isomeric states in the odd-odd nuclei around $^{150}$Pm, having half lives greater or similar to that of the ground state, undergo $\beta$ decay and have very low IT decay probability to the ground state. The presence of these isomers have important effect on the $\beta$ decay branching of the slow neutron capture process (s-process) of nucleo-synthesis. The evidence of the existence of such branch point around $^{148}$Pm and its effect on the s-process path and s-process neutron density is discussed in the work of Lesko et al.~\cite{Pm148}. Hence, the characterization of these isomers close to the branch point is very important also to understand the synthesis of heavy nuclei around the $\beta$ stability line.

The experimental information on the low lying excited states of $^{150}$Pm is limited to a preliminary work based on in-beam spectroscopy using (p,n$\gamma$), ($^3$He,d) and (p,t) reactions~\cite{pm150-prc}. However, the isomers with $\tau _{\frac{1}{2}} \sim sec$ or more can not be detected in an in-beam experiment and hence, the decay measurement becomes very crucial. Identification of these $\beta$ decaying isomers can be done by studying the decay curves for the observed $\gamma$ rays of the daughter and the $\gamma - \gamma$ coincidence measurements involving these decay $\gamma$ transitions. Measurement of $\beta$ decay end-point energies, when measured in coincidence with the daughter $\gamma$ transitions, is also important in order to assign the excitation energy of the isomeric level, especially in absence of any isomeric transition decaying to the ground state of the parent. The measurement of intensities of the $\beta$-feeding gives the information on the {\it logft} values which in turn can be used to guess the spin of the isomeric level and the spin of the excited states of the daughter. Only a single decay spectroscopy work of $^{150}$Pm by Barret et al.,~\cite{barret} exists in literature in which the $\gamma$ rays were detected with a Ge(Li) detector and the coincidence information was obtained by using a well type Ge(Li) detector with the help of summing technique. In this work, the decay curves were not studied for the observed $\gamma$ transitions which can be helpful in verifying their placement in the decay scheme of 2.7 h $^{150}$Pm ground state. This might be the reason for which the identification of any isomeric state in $^{150}$Pm was not possible in this work. Hence, the decay spectroscopy of $^{150}$Pm must be performed with the careful measurement of decay half lives followed by the observed transitions and the $\gamma -\gamma$ coincidence to identify the presence of any long lived isomer in $^{150}$Pm. Also, no measurement exists for the $\beta$ decay end point energies in the decay of $^{150}$Pm to the excited states of $^{150}$Sm~\cite{ensdf-a150}.

In the present work, decay spectroscopy of $^{150}$Pm has been carried out, on the basis of the study of systematcs, to search for the existence of any long-lived isomeric state in this nucleus. The measured half lives followed by the observed transitions and their $\gamma -\gamma$ coincidence relationship indicate the presence of a 2.2(1)h isomeric state in odd-odd $^{150}$Pm for the first time. The $\beta$ decay {\it logft} values have been obtained to guess the spin parity of the levels in $^{150}$Sm daughter nucleus which are fed by the isomer in $^{150}$Pm. The $\beta$ end-point energies corresponding to several decay branches of $^{150g}$Pm $\rightarrow$ $^{150}$Sm decay were measured for the first time by using the $\beta-\gamma$ coincidence technique ~\cite{beta-tumpa}. The energy of the isomeric state has been determined from the measurement of $\beta$ end-point energies corresponding to the isomeric decay. A systematic study of all such isomers existing around $^{150}$Pm has been carried out. Shell Model calculation has been performed by using OXBASH code~\cite{BAB94} that shows the presence of a 5$^-$ isomer in the level scheme of $^{150}$Pm. The calculation also indicates the possibility of $\beta$ decay from this isomer with a highly hindered electromagnetic transition probability.

\section{Experimental Details}
\label{expt}
The excited states of $^{150}$Pm was populated by the $^{150}$Nd (p, n)$^{150}$Pm reaction using 8.0 MeV proton beams provided by K = 130 AVF cyclotron at Variable Energy Cyclotron Centre~(VECC), Kolkata. The beam energy was chosen from the measured excitation function for the p~$+$~$^{150}$Nd reaction~\cite{pm150-cs} so as to have maximum cross section for the (p,n) channel. The $^{150}$Nd target was prepared  by electro-deposition technique, starting from commercially available 97.65$\%$ enriched powdered oxide sample~(Nd$_2$O$_3$), on a 0.3 mil thick Aluminium (Al) foil. The isotopic impurities in the target material consisted of 0.50$\%$ $^{142}$Nd, 0.31$\%$ $^{143}$Nd, 0.68$\%$ $^{144}$Nd, 0.23$\%$ $^{145}$Nd, 0.47$\%$ $^{146}$Nd and 0.26$\%$ $^{148}$Nd, as per the data sheet provided by the supplier. The thickness of the targets used in the experiment was $\sim$900 $\mu$g/cm$^2$. Several targets were irradiated following the stacked foil technique~\cite{pm150-cs}. The irradiated targets were subsequently counted with different configurations of Ge detectors as per the requirement of the decay spectroscopy measurements as discussed below.

In the present work, the decay $\gamma$ transitions have been detected using the VENUS~(\underline{VE}CC Array for \underline{NU}clear \underline{S}pectroscopy) array of six Compton suppressed Clover HPGe (High Purity Ge) detectors~\cite{soumik-venus} which were placed at the angles of 30$^{\circ}$, 90$^{\circ}$, 180$^{\circ}$, 260$^{\circ}$ and 310$^{\circ}$ w.r.t one of the detectors taken as a reference of 0$^{\circ}$ and at a distance of 18 cm from the target. In order to measure the decay half lives, the $\gamma$ decay was followed in singles mode over a period of $\sim$10h where each counting was continued for a duration of 10 minutes. It was also ensured that the detector dead time is kept below 5$\%$ during counting. In the $\gamma-\gamma$ coincidence measurement, conventional two fold coincidence logic was used for gathering the data with the VENUS array.

The $\beta$ decay endpoint energies were determined from $\beta-\gamma$ coincidence measurement~\cite{beta-tumpa} by using two Low Energy Photon Spectrometer (LEPS), having 300~$\mu$m Be window, in conjunction with four Clovers of the VENUS array. During this measurement, only the bare target has been used by placing it in such a way that the irradiated target material faces the LEPS detectors and ensures the efficient emission of $\beta$ particles from the target material. Each LEPS detector is made up of a planar HPGe crystal which is electrically separated in four segments providing four individual signals. Data were gathered in two different modes of (i) $\gamma -(\beta + \gamma)$ and (ii) $\gamma - \gamma$ establishing two fold coincidence among the set of four Clover detectors and the eight segments of the two LEPS detectors, respectively. In this measurement, the $\gamma - \gamma$ coincidence data was taken by placing Al blocks of 10 mm thickness in between the detector and the target in order to completely stop the $\beta$ particles from reaching the detectors as and when required. For the $\beta-\gamma$ measurements, the target to detector distance was kept as 10.5 cm.

The pulse processing for the present work was performed by using sixteen channel Mesytec high resolution Shaper and Amplifier module and VME DAQ system LAMPS~\cite{lamps} with the 32 bit Mesytec ADC~(MADC-32). All the decay measurements have been carried out after a cooling time of $\sim$2~h or more in order to ensure that all the short lived activities produced during irradiation have decayed out. Measurements of absolute efficiencies of all the detectors were performed using standard technique with the $^{152}$Eu and $^{133}$Ba sources of known activity. All the data were gathered in zero suppressed LIST mode and the offline analysis were performed using the LAMPS and RADWARE~\cite{radware} analysis packages.

\section{Data Analysis and Results}
\label{da}

The data analysis was performed in order to (i) determine the decay half lives followed by the observed $\gamma$ transitions, (ii) study the $\gamma-\gamma$ coincidences, (iii) measurement of $\gamma$ intensities $\&$ calculation of {\it logft} values  and (iii) determine the $\beta$ decay end point energies. The details of data analysis and the results obtained from the present work have been discussed in the following subsections.

\subsection{Measurement of Decay Half Lives}
\label{decay}

\begin{figure}
  \begin{center}
  \hskip -1.0cm
  \includegraphics[width=\columnwidth]{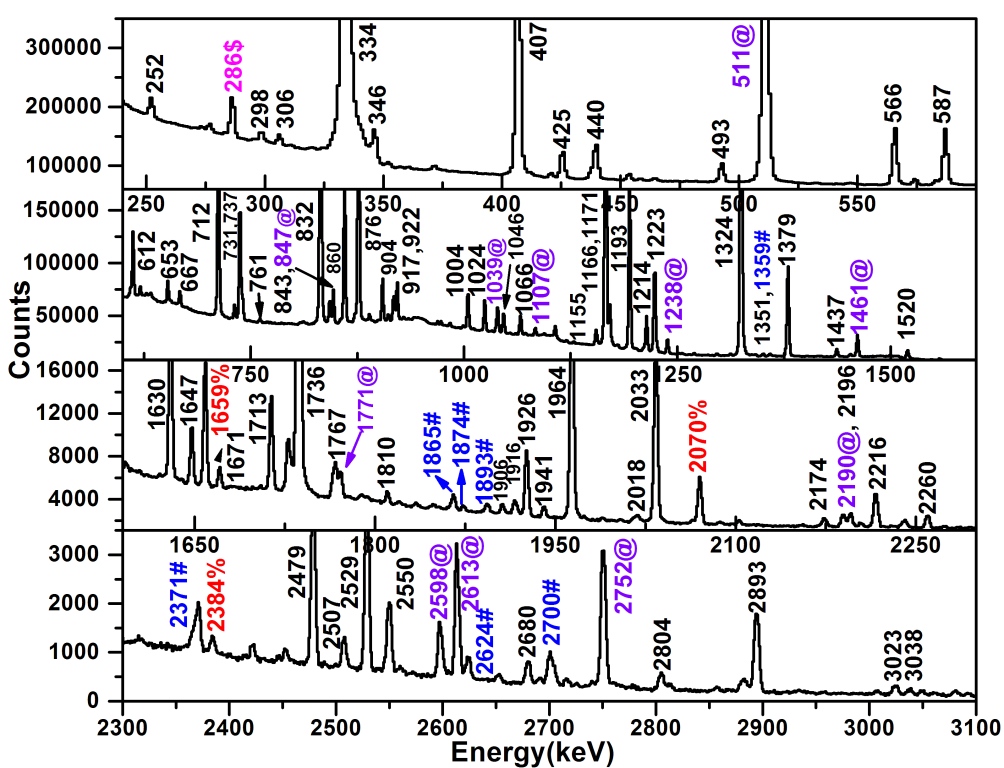}
  \caption{(Color Online)~The singles spectrum showing all the transitions observed from the decay measurement. The transitions with no marking (in black) are already placed in the level scheme of $^{150}$Sm from the decay of $^{150}$Pm. The transitions marked with $\#$ (in blue) were, although assigned to $^{150}$Sm~\cite{ensdf-a150}, but not placed in the level scheme in the earlier experiments. The transitions marked with $\%$ (in red) are the new transitions observed for the first time in the present work. The 286 keV transition,  marked with $\$$ (in pink), is identified to be arising from the decay of $^{149}$Pm~($\tau _{1/2}$~=~53.08(5)~h). The (violet) transitions marked with @ follow longer half lives and have been identified to be coming from background or the contaminations in the Al backing foil.}
  \vspace{-8mm}
  \label{fig1-totspec}
  \end{center}
  \end{figure}

The $\gamma$ transitions obtained from the singles mode counting are shown in Fig.~\ref{fig1-totspec} and the decay curves exhibited by all of them were studied in order to identify the origin of each transition. This was performed by using the standard techniques where the background subtracted areas of a particular full energy peak were studied as a function of time. Most of the observed transitions belong to the level scheme of $^{150}$Sm, the $\beta ^-$ decay daughter nucleus of $^{150}$Pm which is produced via (p,n) reaction. The 286 keV $\gamma$ transition which is known in the level scheme of $^{149}$Sm, produced via $\beta$ decay of the (p,2n) reaction channel, was also found to be present in the spectrum. No $\gamma$ transition has been identified which are originated from other Nd isotopes present in the enriched $^{150}$Nd target. Three new transitions were found in the singles data and the placement of these transitions could be confirmed in the level scheme of $^{150}$Sm by following decay half lives and $\gamma -\gamma$ coincidence analysis. However, some of the transitions are observed to be decaying with much longer half lives and these were identified to be originating from either the impurities in the Al backing foil or from background.

\begin{figure}
  \begin{center}
  \includegraphics[width=\columnwidth]{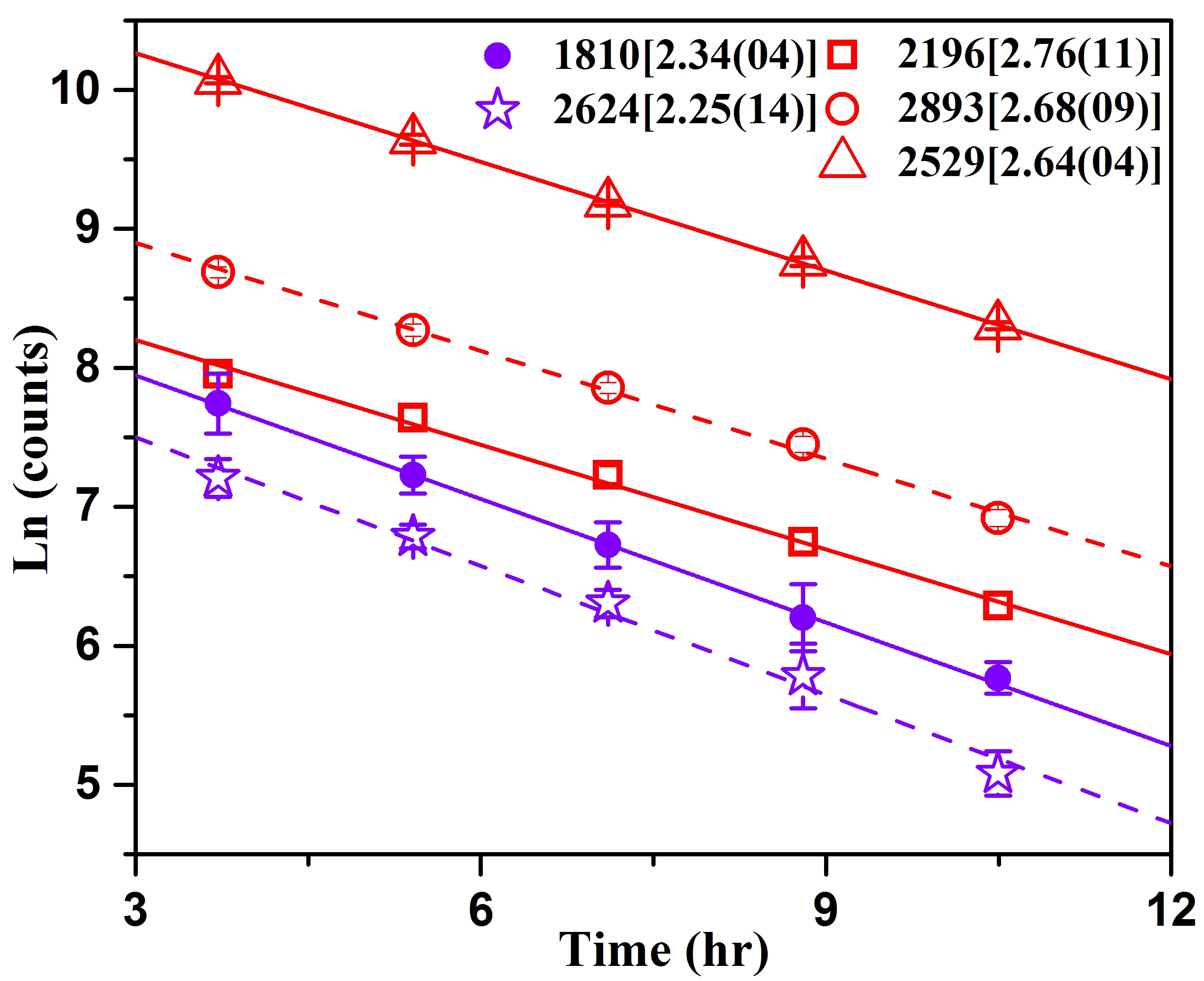}
  \caption{(Color Online) The decay curves are shown for two groups of transitions, corresponding to two different half lives. The data points could be fitted with straight lines having different slopes and are shown. Different symbols correspond to different transition energies which are indicated along with the obtained half life values.}
  \label{fig2-halflife}
  \end{center}
  \end{figure}

Excluding the transitions which were identified as coming from sources other than $^{150}$Sm, two main groups of transitions were observed. One of these two groups is observed to decay with half life similar to the half life for $^{150}$Pm ground state~($\sim$2.7~h) and the other group exhibit decay curves with half life $\sim$2.2~h which is less than the half life of the $^{150}$Pm ground state. Fig.~\ref{fig2-halflife} shows the representative experimental decay curves for some of the transitions belonging to the said two groups.

\begin{figure}
  \begin{center}
  \includegraphics[width=\columnwidth]{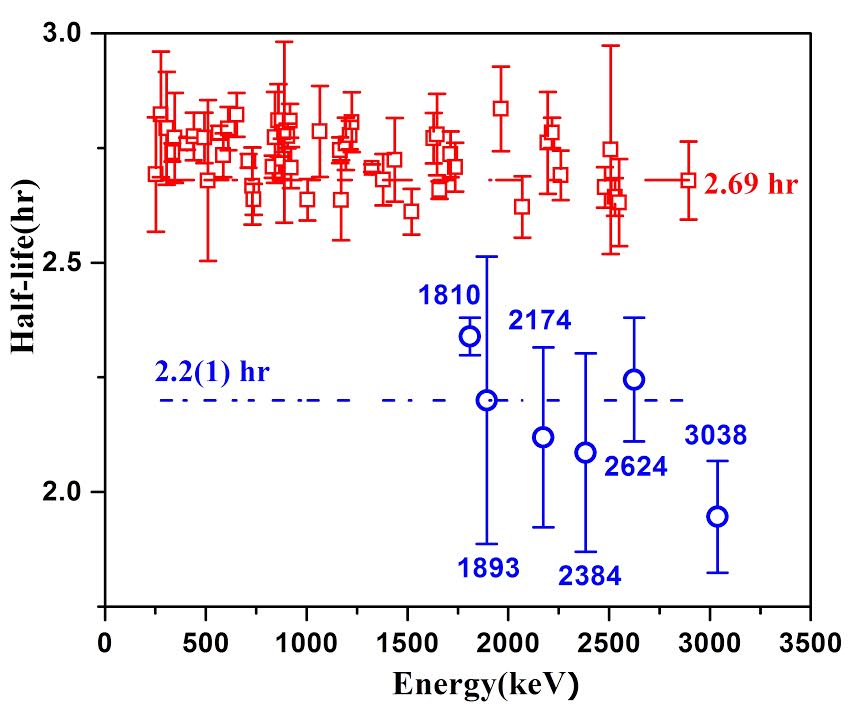}
  \caption{(Color Online) The half life values obtained by following the yields of the observed transitions are shown as a function of energy. The data points corresponding to different half lives are indicated with different colours and symbols.}
  \label{fig3-halflife-all}
  \end{center}
  \end{figure}

The observation has been summarised by showing Fig.~\ref{fig3-halflife-all} where the half life values followed by all the observed transitions of $^{150}$Sm have been plotted as a function of energy. In the present work, six transitions, viz., 1810, 1893, 2174, 2384, 2624 and 3038 keV transitions have been observed to exhibit decay curves with half lives $\sim$ 2.2~h. Out of these, all the five transitions, viz., 1810, 1893, 2174, 2624 and 3038 keV, are known in the ENSDF database of $^{150}$Sm~\cite{ensdf-a150}. However, the 1893 and 2624 keV transitions were not placed in the level scheme of $^{150}$Sm, although these two transitions were known from the decay of $^{150}$Pm~\cite{barret}. The 2384 keV transition is observed for the first time in the present work but follows half life different than the ground state of $^{150}$Pm. In the present work, all these six transitions were confirmed to be present in $^{150}$Sm from the analysis of $\gamma -\gamma$ coincidences as described in the following subsection. From the observation of two different half lives followed by the $\gamma$ transitions belonging to the level scheme of $^{150}$Sm, which can only be produced from the $\beta$ decay of $^{150}$Pm, the present work proposes the existence of a $\beta$ decaying isomer in the level scheme of $^{150}$Pm. The half life for the proposed $\beta$ decaying isomer in $^{150}$Pm was determined from the weighted average value of the half lives followed by these six transitions and it comes out to be 2.2(1)~h.

\subsection{Measurement of $\gamma-\gamma$ coincidences}
\label{gg}

\begin{figure}
  \begin{center}
  \includegraphics[width=\columnwidth]{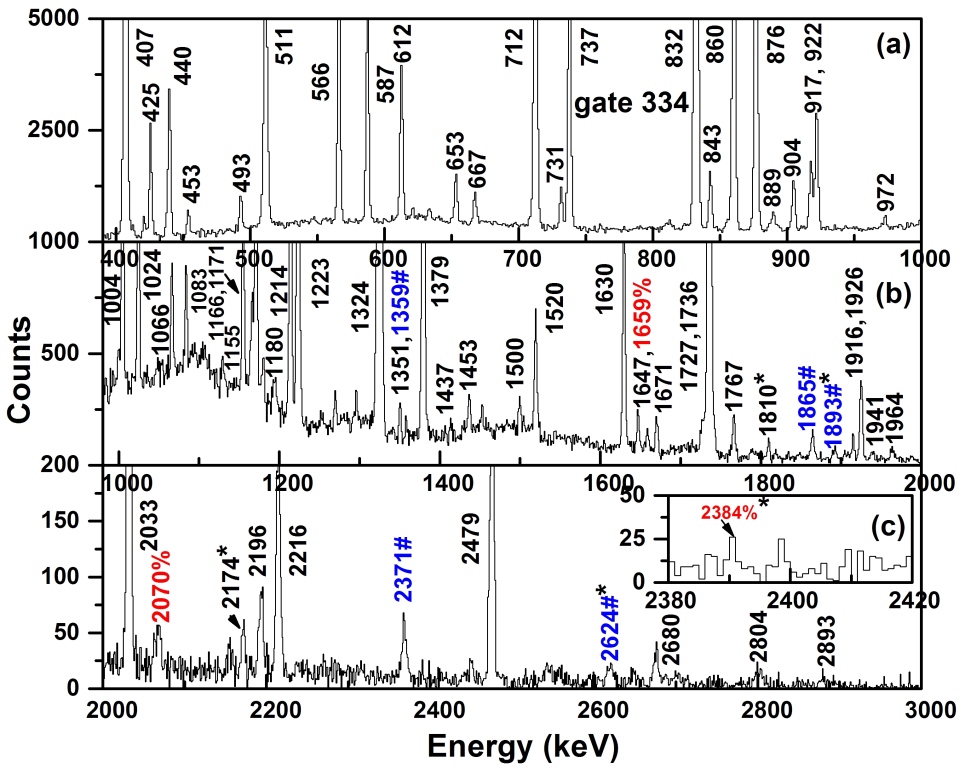}
  \caption{The 334 keV (2$^+$ $\rightarrow$ 0$^+$ transition in $^{150}$Sm) gated spectra showing the transitions belonging to $^{150}$Sm nucleus. The transitions showing half lives $\sim$2.2~h has been marked with stars. The peaks marked with colours and with symbols other than * has the same representation as in Fig.~\ref{fig1-totspec}.}
  \label{fig8-gate334}
  \end{center}
  \end{figure}

\begin{figure}
  \begin{center}
  \includegraphics[width=\columnwidth]{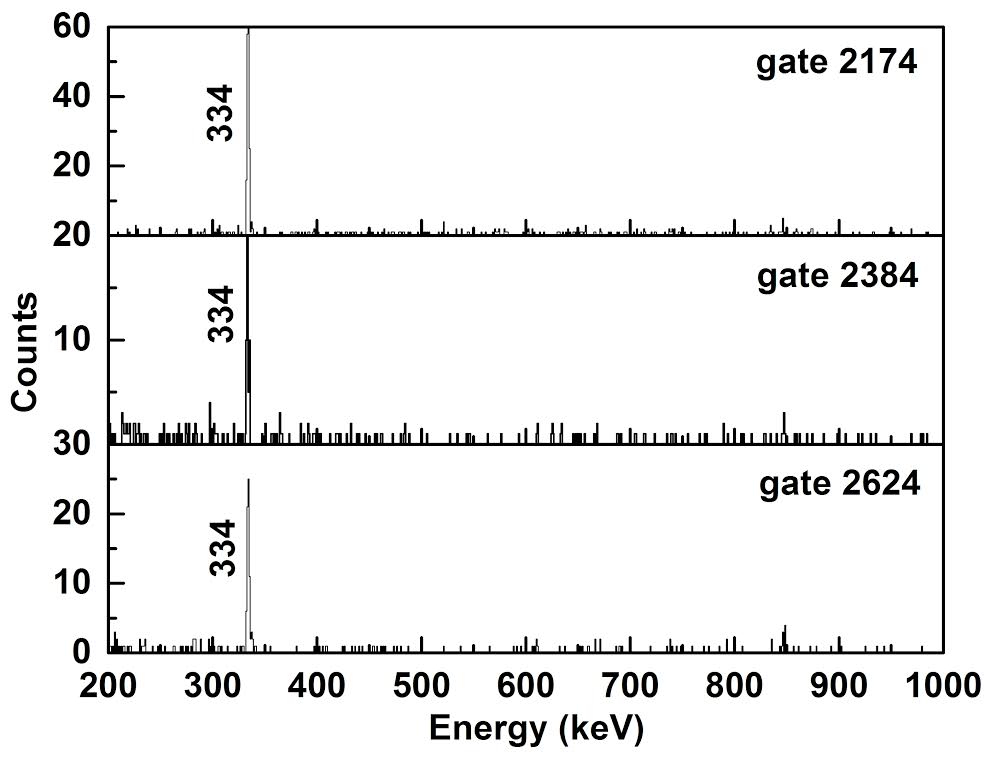}
  \caption{The gated spectra for three higher energy transitions showing half lives $\sim$2.2~h.}
  \label{fig8a-gate22}
  \end{center}
  \end{figure}

\begin{figure}
  \begin{center}
  \includegraphics[width=\columnwidth]{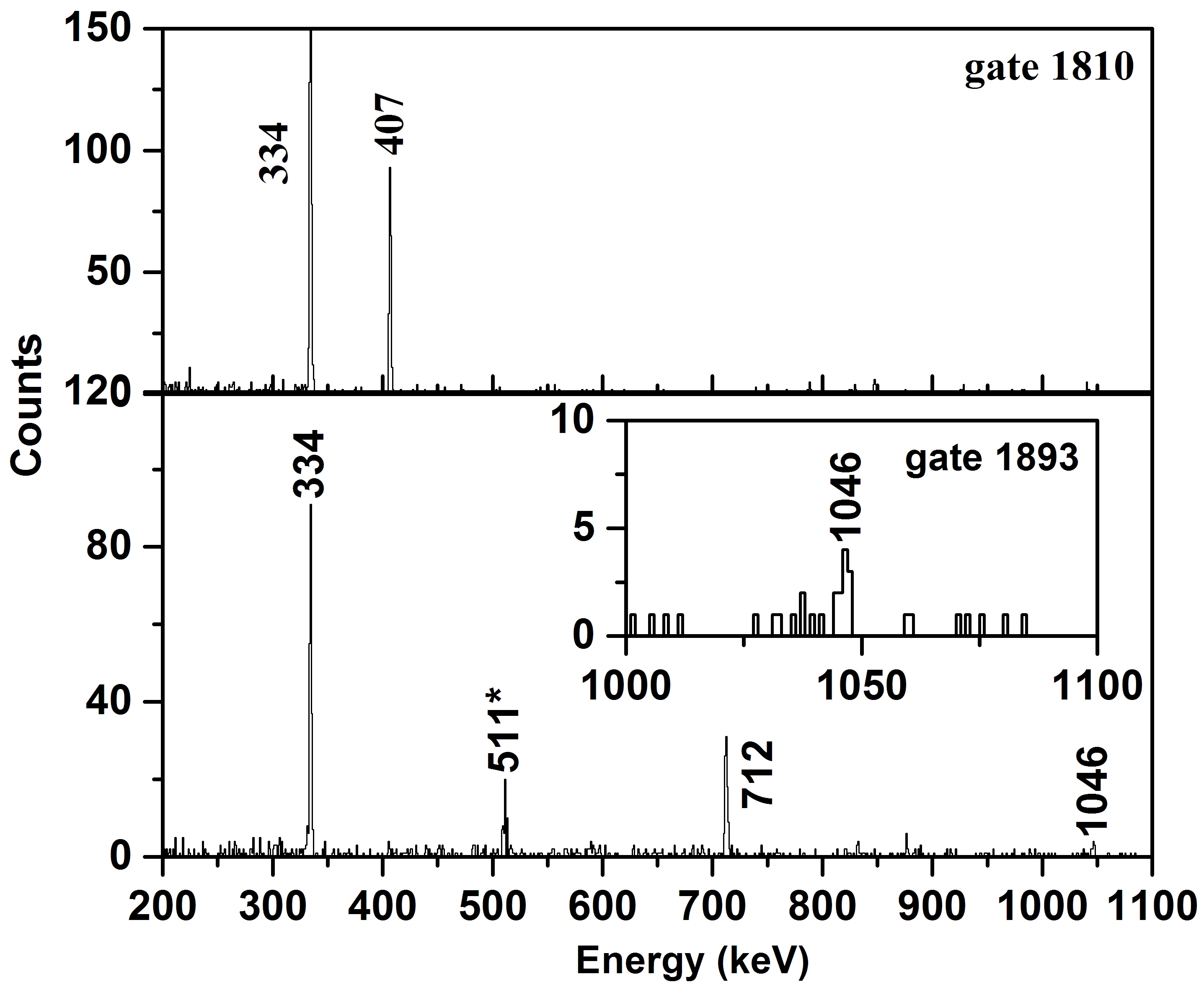}
  \caption{The gated spectra for 1810 and 1893 keV transitions. The 511 keV transition is marked with $\star$.}
  \label{fig8b-gate22}
  \end{center}
  \end{figure}

The $\gamma -\gamma$ coincidence analysis was performed only for those transitions which are relevant for the development of the decay scheme of the observed isomer and the same has been described below. The $\gamma -\gamma$ coincidence information regarding all other transitions belonging to the decay scheme of $^{150g}$Pm and observed in the present experiment may be reported in future. Out of the six transitions following half lives $\sim$2.2h, the 3038 keV transition is known to be directly feeding the ground state of $^{150}$Sm. The remaining five transitions, viz., 1810, 1893, 2174, 2384 and 2624 keV, could be found in the 334 keV gated spectrum, as shown in Fig.~\ref{fig8-gate334}.
 The gated spectra for these five transitions, viz., 1810, 1893, 2174, 2384 and 2624 keV, have been shown in Fig.~\ref{fig8a-gate22} and Fig.~\ref{fig8b-gate22}. The three higher energy transitions, viz., 2174, 2384 and 2624 keV, are found to be in coincidence only with the 334 keV(2$^+$ $\rightarrow$ 0$^+$) transition of $^{150}$Sm. The lower energy transitions, viz., 1810 and 1893 keV, display coincidence with other higher lying transitions of $^{150}$Sm as well.

The 1893 keV transition was known in the ENSDF database but was not assigned to the level scheme of $^{150}$Sm~\cite{ensdf-a150}. In the present work, the transition has been confirmed to be present in the level scheme of $^{150}$Sm from the $\gamma -\gamma$ coincidence analysis, as shown in Fig.~\ref{fig8b-gate22}. From the coincidence data, the energy of the excited level comes out to be 2939 keV that de-excites by the 1893 keV transition. A 2937 keV level is known from the (n,$\gamma$), (p,t) and (d,p) reactions, however, no $\gamma$ transition was known to decay from this level. Also, the energy of this level is uncertain by 20 keV as reported in ENSDF database~\cite{ensdf-a150}. The 2939 keV level proposed in the present work could be same as the 2937 keV level reported from earlier works.

The 2624 keV transition is known from the decay of $^{150}$Pm~\cite{barret} and in the present work, it could be placed in the level scheme of $^{150}$Sm from the $\gamma -\gamma$ coincidence analysis, as shown in Fig.~\ref{fig8a-gate22}. It is found to be decaying from a 2958 keV excited state in $^{150}$Sm as per the observed coincidence with the 334 keV transition. This 2958 keV level has been assigned in the level scheme of $^{150}$Sm for the first time in the present work.

The 1810 keV transition is known to be decaying from the 2550.6 keV level, as per the $\beta$ decay work~\cite{barret}. The $\gamma-\gamma$ coincidence analysis, as shown in Fig.~\ref{fig8b-gate22}, confirms that the 1810 keV transition is in coincidence with 334 and 407 keV transitions. The 2550.6 keV level also decays via two more transitions, viz., 2550.5 and 2216.5 keV. However, these two transitions were found to follow half lives close to the half life of the $^{150}$Pm ground state and, thus, can not be originated from the proposed isomeric level. Moreover, an accurate energy measurement of the involved transitions (1809.9, 334.3 and 406.8 keV) suggests that the 1810 keV $\gamma$ ray is decaying from a 2551 keV level. The similar energy measurement for the earlier known 2550.5 and 2216.5 keV transitions yields the values of 2549.5 keV and 2216.1 keV respectively. Hence, it is proposed that there might be two close lying levels existing in $^{150}$Sm, one at 2550 keV, decaying the 2549.5 and 2216.1 keV transitions and the other at 2551 keV de-exciting via 1810 keV transition. In the present work, the latter level is proposed to be fed by the $\sim$2.2 h isomeric state in $^{150}$Pm.

The 2174 keV transition is decaying from the 2507.3 keV level from which four more transitions of energy 2507, 1437, 1341 and 848 keV are known to be decaying out. In the present work, the 848 keV has been found to follow longer half life and thus, confirmed to be originating from contamination in the Al backing. The 1341 keV transition was not observed in the present work. The two remaining transitions of energy 2507 and 1437 keV show decay curves with half life close to the half life of the ground state of $^{150}$Pm. In the ENSDF database of $^{150}$Sm, two levels are known at around 2507 keV. One of them is 2507.27 keV,(1$^-$, 2$^+$) level assigned from the $\beta$ decay work and the other is 2507.5 keV, 3$^+$, 4$^+$ level which was observed from the (n, $\gamma$) reaction. It is, thus, conjectured that the 2507.5 keV level, having higher spin value (3$^+$, 4$^+$), is fed by the 2.2~h isomer in $^{150}$Pm and decays by the 2174 keV $\gamma$ transition. This was based on the observation of difference in half life followed by the 2174 keV transition compared to the other two transitions, viz., 2507 and 1437 keV, decaying from the 2507.3 keV level and the $\gamma -\gamma$ coincidence information, as obtained from Fig.~\ref{fig8a-gate22}.

The 2384 keV transition is observed for the first time in the present work and shows coincidence with the 334 keV transition~(cf.~Fig.~\ref{fig8a-gate22}). Following its coincidence relationships, this $\gamma$ ray has been placed as decaying from the 2718 keV level. One 2715(4) keV, 3$^-$ level is known in $^{150}$Sm from the (p,p$^{\prime}$) and (d,d$^{\prime}$) reactions~\cite{ensdf-a150}. However, no $\gamma$ transition is known to be decaying from this level. In the present work, it is proposed that the newly observed 2718 keV level could be same as the 2715(4) keV level reported already in ENSDF database.

 The 3038 keV level is also known in the level scheme of $^{150}$Sm from which the 3038 keV transition is known to be decaying directly to the ground state of $^{150}$Sm. As mentioned above, this 3038 keV transition has been observed in the present work and follows a half life $\sim$2.2~h.

\begin{figure}
  \begin{center}
  \hskip -1.0cm
  \includegraphics[width=\columnwidth]{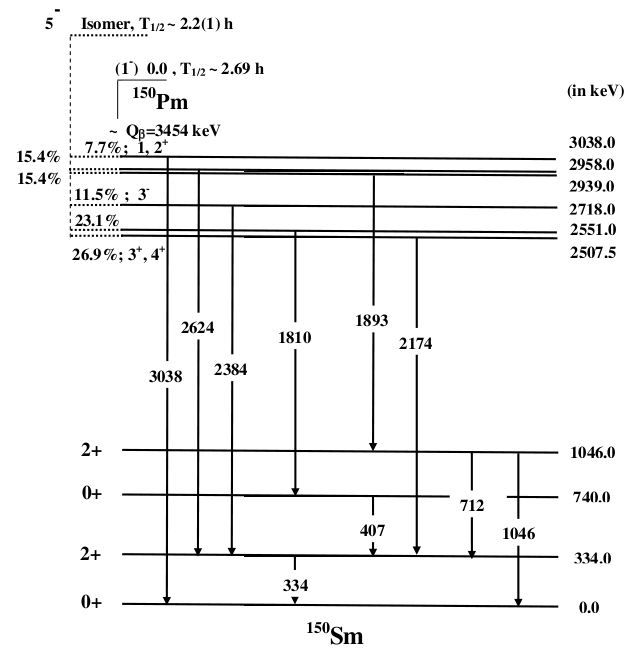}
  \caption{The decay scheme obtained in the present work for the newly identified 2.2(1)~h isomeric state in $^{150}$Pm. The ground state Q value is indicated. The spin parity of the isomer is suggested from shell model calculation (cf. Fig.~\ref{shmod}).}
  \label{fig9-isomer2.2}
  \end{center}
  \end{figure}

From the study of coincidence relationships among transitions following a half life $\sim$2.2~h, as described above, the decay scheme for the proposed $\beta$ decaying isomeric state in $^{150}$Pm is developed and has been shown in Fig.~\ref{fig9-isomer2.2}. A spin  parity of 5$^-$ is predicted for this isomer from the shell model calculation (cf.~Section~\ref{shell}) and the same has been indicated in the figure. From the present work, no transition was found that could be identified as decaying from the said 5$^-$ isomeric level to the ground state of $^{150}$Pm which has been suggested to have a j$^{\pi}$ value of 2$^-$ in the work of D. Bucurescu et al~\cite{pm150-prc} and also from the shell model calculation (cf.~Section~\ref{shell}). However, the present work could also not rule out the existence of any such $\gamma$ decay. It can be conjectured that the electromagnetic decay from the proposed isomeric level might be highly hindered as it is possibly a low energy M3 or E4 $\gamma$ transition.

\subsection{Measurement of $\gamma$ intensities and calculation of {\it logft} values}
\label{i-logft}

The intensity of $\beta$ feedings from the proposed isomeric state has been obtained from the intensities of the involved $\gamma$ transitions, measured in the present work. For this purpose, the detector efficiency was measured with the standard $^{133}$Ba and $^{152}$Eu sources, which has maximum $\gamma$ energy up to 1408 keV. Hence, for determining the efficiencies at higher energies upto 3038 keV, a linear extrapolation of the efficiency curve was made. Table~\ref{tab-int} shows the intensity values for all the transitions in $^{150}$Sm that are emitted following the decay of the proposed isomeric states in $^{150}$Pm along with other relevant information. The intensities of the transitions have been obtained from the singles data and by considering the intensity of the 334 keV (2$^+$ $\rightarrow$ 0$^+$) transition as 100$\%$. Considering the total observed $\beta$ decay from this particular isomer as 100$\%$, the percentage of $\beta$-feeding to the corresponding levels of $^{150}$Sm has been calculated from the intensity of the decaying $\gamma$ transitions and are shown in Fig.~\ref{fig9-isomer2.2}. The {\it logft} values have been calculated for all these $\beta$ decay branches by considering the energy of the isomeric level in $^{150}$Pm at the following three different values - 33.0 keV, as indicated by the shell model calculation (cf.~section~\ref{shell}), 454(284) keV, as indicated from the $\beta$ decay end point energy measurement(cf.~section~\ref{beta}) and  150 keV which is close to the isomeric levels known in $^{148}$Pm and $^{152}$Pm nuclei. The calculation of {\it logft} values have been performed by using the {\it logft} calculation program available in the NNDC database~\cite{ensdf-logft}.
\begin{table*}
 \begin{center}
 \begin{small}
 \caption{Details of the excited levels and $\gamma$ $\&$ $\beta$ transitions relevant to the decay of isomeric level in $^{150}$Pm.}
 \label{tab-int}
 \begin{tabular}{|ccc|ccc|c|c|c|ccc|}
 \hline
 &&&&&&&&&&&\\
\multicolumn{3}{|c|}{The isomeric level in $^{150}$Pm}&\multicolumn{3}{c|}{Levels in $^{150}$Sm}&Decay E$_{\gamma}$&I$_{\gamma}$&I$_{\beta}$\footnotemark&\multicolumn{3}{c|}{logft with}\\
 &&&&&&&&&\multicolumn{3}{c|}{diff E$_x$}\\
 Ex. Energy(E$_x$)&J$^{\pi}$&Lifetime&Ex. Energy&\multicolumn{2}{c|}{J$^{\pi}$}&&&&\multicolumn{3}{c|}{of $^{150m}$Pm}\\
 &&&&&&&&&&&\\
 (keV)&&(h)&(keV)&(ENSDF~\cite{ensdf-a150})&(Proposed)&(keV)&$\%$ of&($\%$)&33&150&454\\
 &&&&&&&334 keV&&keV&keV&keV\\
 &&&&&&&(2$^+$ $\rightarrow$ 0$^+$)&&&&\\
 &&&&&&&&&&&\\
 \hline
 &&&&&&&&&&&\\
 \multicolumn{3}{|l|}{Expt:}&3038.0&1,2$^+$&(4$^{\pm}$,5$^{\pm}$,6$^{\pm}$)&3038.0&0.020(4)&7.7&5.5&5.8&6.5\\
 &&&&&&&&&&&\\
  453(284)&-&2.2(1)&2958.0&-&(4$^{\pm}$,5$^{\pm}$,6$^{\pm}$)&2624.0&0.06(1)&15.4&5.4&5.7&6.3\\
 &&&&&&&&&&&\\
 \multicolumn{3}{|l|}{Shell Model:}&2939.0&-&(4$^{\pm}$,5$^{\pm}$,6$^{\pm}$)&1893.0&0.06(1)&15.4&5.5&5.8&6.4\\
 &&&&&&&&&&&\\
 33&5$^-$&-&2718.0&3$^-$&(4$^{\pm}$,5$^{\pm}$,6$^{\pm}$)&2384.0&0.05(1)&11.5&6.1&6.3&6.8\\
 &&&&&&&&&&&\\
 \multicolumn{3}{|l|}{Systematics:}&2551.0&-&(4$^{\pm}$,5$^{\pm}$,6$^{\pm}$)&1810.0&0.05(1)&23.1&6.1&6.3&6.7\\
 &&&&&&&&&&&\\
 $\sim$150&-&-&2507.5&3$^+$,4$^+$&(4$^+$)&2174.0&0.05(1)&26.9&6.1&6.3&6.7\\
 &&&&&&&&&&&\\
 \hline
 \end{tabular}
 \end{small}
 \footnotetext{The intensities of $\beta$ branching was calculated by considering the sum of all the observed decay of $^{150m}$Pm as 100$\%$}.
 \end{center}
 \end{table*}

The {\it logft} values obtained for the 2939.0, 2958.0 and 3038.0 keV levels comes in the range of 5.4 to 6.5 for three types of excitation energies of the excited states. This range of values of {\it logft} has been assigned mostly for {\it allowed} $\Delta$J = 0,1 or {\it first forbidden} $\Delta$J = 0,1 transitions~\cite{tuli} and this leads to conjecture the J$^{\pi}$ values of these levels as (4$^{\pm}$,5$^{\pm}$,6$^{\pm}$). Hence, the assigned spin parity for the 3038 keV level, as obtained from the earlier beta decay measurement~\cite{ensdf-a150}, may be an artifact of consideration of the ground state beta decay of $^{150}$Pm to be responsible for the population of this level.

The {\it logft} value obtained for the 2507.5, 2551 and 2718 keV levels comes about 6.1 to 6.8 as shown in table~\ref{tab-int}. This value also corresponds to {\it allowed} and {\it first forbidden unique} transitions~\cite{tuli}. Out of the said three levels, the J$^{\pi}$ value for the 2507.5 keV level is known to be (3$^+$, 4$^+$) from the (n,$\gamma$) work~\cite{ensdf-a150}. Consideration of the 3$^+$ spin for this level demands the beta decay feeding from the isomeric state to be of {\it first forbidden unique} type. In the present work, the spin parity of 4$^+$ is proposed for this level based on the extracted beta decay $\it{logft}$ value and a 5$^-$ spin for the isomeric level. Also an assignment of 3$^-$ to the 2718 keV level (previously 2715 keV observed from (p, p$^{\prime}$) reaction), that claims a {\it second forbidden non unique} type beta decay, is not appropriate from the observed $\it{logft}$ value. Hence, either these two levels are different and 2718 keV is a new level observed in the present work or the spin parity assignment to this level has to be different from 3$^-$. In the present work, the spin parity of (4$^{\pm}$,5$^{\pm}$,6$^{\pm}$) has, thus, been assigned to the 2718 keV level. The spin parity of 2551 keV level is also proposed to be (4$^{\pm}$,5$^{\pm}$,6$^{\pm}$) as per the observed beta decay $\it{logft}$ value.

Hence, an accurate spin parity measurement is envisaged for all these particular levels in $^{150}$Sm which was not possible in the present work.

\subsection{Measurement of $\beta$ decay End Point Energy}
\label{beta}

The $\beta$ decay end point energies have been measured for several decay branches of $^{150}$Pm to $^{150}$Sm, following the techniques described in Ref.~\cite{beta-tumpa}. In order to generate the $\gamma - (\beta + \gamma$) and $\gamma -\gamma$ coincidence information, two RADWARE compatible matrices were made from the coincidence data obtained with two LEPS and four Clover HPGe detectors, in two different configurations as described in section~\ref{expt}. These LEPS events were compressed to 10, 20 and 40 keV/channel for the generation of $\beta$ decay spectrum as and when required and with appropriate $\gamma$ energy gates from the Clover events.

\begin{figure}
  \begin{center}
  \hskip -1.0cm
  \includegraphics[width=\columnwidth]{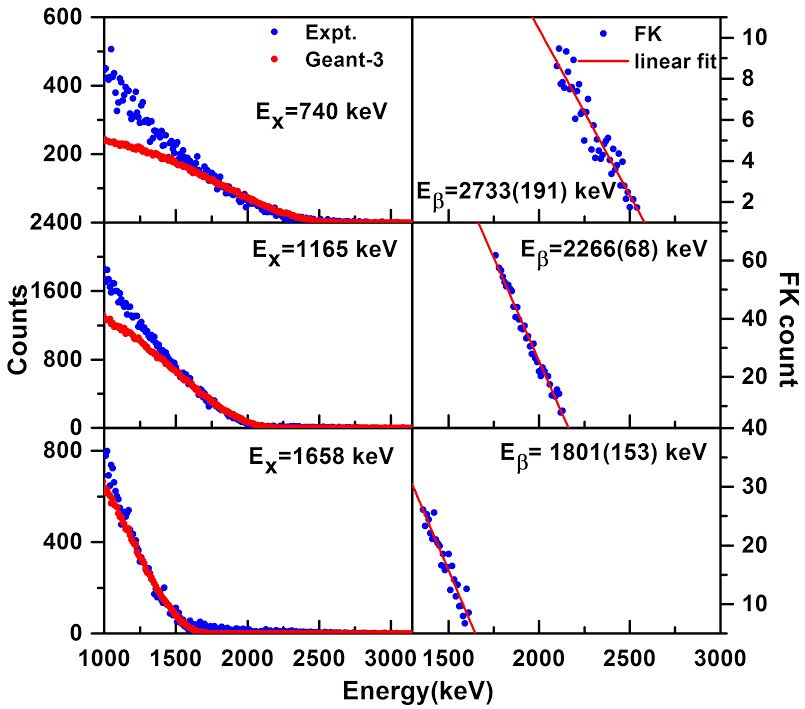}
  \caption{(Color Online) The representative $\beta$ spectra obtained for some of the decay branches of $^{150g}$Pm $\rightarrow$ $^{150}$Sm. The experimental data points have been shown with blue circles and the Red line shows the results obtained from Geant3 simulation. The FK plots are shown with the obtained end point energies.}
  \label{fig10-beta}
  \end{center}
  \end{figure}

The representative $\beta$-spectra obtained in the present work for the decay branches of $^{150}$Pm to $^{150}$Sm have been shown in Fig.~\ref{fig10-beta}. The corresponding $\beta$-spectra have also been obtained by using Monte Carlo simulation with the Geant3 simulation package~\cite{geant3} with one million events and the exact geometrical configuration. The results from the simulation has been normalized to match the experimental data points based on visual estimation, with an emphasis in the energy range of 0.8E$_{\beta}$ to E$_{\beta}$ and is plotted with the corresponding experimental results as shown in Fig.~\ref{fig10-beta}. The deviation of the experimental results from the simulation at the low energy side corroborates to the effects of background appearing from the $\beta$-Compton and Compton-Compton coincidences that could not be subtracted during generation of $\beta$ spectrum as has been described in Ref.~\cite{beta-tumpa}. The Fermi-Kurie~(FK) plots were generated from the experimentally obtained $\beta$ spectrum by using the built in routine {\it `FK-Energy'} included in the spectrometer code LISE~\cite{lise,kantele}. The FK plots corresponding to the different $\beta$ decay branches have been shown in Fig.~\ref{fig10-beta} along with the $\beta$ spectrum. The $\beta$ end point energies have been obtained by fitting these plots and the results have been tabulated in Table~\ref{Table-beta} corresponding to different decay branches. The required corrections were made for the attenuation in the Be window of the LEPS detectors and both the corrected and uncorrected energies are mentioned in Table~\ref{Table-beta}. The $\gamma$ energy has also been indicated in the table for each $\beta$ branch which corresponds to the gating transition that was used for generating the $\beta$ spectrum.
\begin{table*}
\begin{center}
\caption{The $\beta$ decay end point energies obtained in the present work with relevant details, as described in text.}
\begin{small}
\label{Table-beta}
\begin{tabular}{|c|cc|c|c|ccccc|}
\hline
&&&&&&&&&\\
Decaying&\multicolumn{2}{c|}{Final Level}&Branching&Gating transition&\multicolumn{5}{c|}{$\beta$ Energy}\\
Nucleus&&&&&&&&&\\
&Energy&J$^{\pi}$&&&(Q$_{\beta}$ - E$_x$)&Lit.&\multicolumn{2}{c}{F-K Anal.}&GEANT3\\
(Q$_{\beta}$)&(E$_x$)&&&&&&(uncorr)&(corr)&and\\
&&&&&&&&&$\chi ^2$ Anal.\\
&(keV)&&($\%$)&(keV)&(keV)&(keV)&(keV)&(keV)&(keV)\\
&&&&&&&&&\\
\hline
&&&&&&&&&\\
$^{150g}$Pm&334.0&2$^+$&$\le$10&334&3120.0&-&2957$\pm$73&3046$\pm$73&3134$\pm$34\\
3454(20) keV&&&&&&&&&\\
&740.4&0$^+$&1.5&407&2713.6&-&2645$\pm$191&2733$\pm$191&2673$\pm$65\\
&&&&&&&&&\\
&1046.1&2$^+$&1.3&712&2407.9&-&2429$\pm$219&2517$\pm$219&2407$\pm$102\\
&&&&&&&&&\\
&1165.8&1$^-$&25.9&1165+832&2288.2&-&2179$\pm$68&2266$\pm$68&2330$\pm$37\\
&&&&&&&&&\\
&1658.4&2$^{(-)}$&19.4&1324&1795.6&-&1715$\pm$153&1801$\pm$153&1826$\pm$28\\
&&&&&&&&&\\
&1713.5&1&3.5&1379&1740.7&-&1781$\pm$105&1868$\pm$105&1761$\pm$47\\
&&&&&&&&&\\
&1963.7&1$^-$&5.5&1223+1964+1630+917&1490.3&-&1529$\pm$193&1615$\pm$193&1462$\pm$40\\
&&&&&&&&&\\
&2070.3&2$^{(-)}$&17.5&1736+876&1383.7&-&1478$\pm$180&1555$\pm$180&1424$\pm$23\\
&&&&&&&&&\\
&2259.9&(1$^-$)&3.26&1214+1004+1066+1926+1519&1194.1&-&1509$\pm$115&1594$\pm$115&-\\
&&&&&&&&&\\
&2367.4&(3$^+$)&1.04&2033&1086.6&-&1331$\pm$221&1417$\pm$221&1068$\pm$67\\
&&&&&&&&&\\
\hline
&&&&&&&&&\\
$^{150m}$Pm\footnotemark&2551.0&(4$^{\pm}$,5$^{\pm}$,6$^{\pm}$)&23.1&1810&903.4&-&1270$\pm$284&1356$\pm$284&-\\
(3454 keV)&&&&&&&&&\\
&&&&&&&&&\\
\hline
\end{tabular}
\end{small}
\footnotetext{For the isomeric state, the branching has been considered from Fig.~\ref{fig9-isomer2.2} of section~\ref{i-logft}. The Q value is mentioned for ground state decay.}
\end{center}
\end{table*}
\begin{figure}
  \begin{center}
  \hskip -1.0cm
  \includegraphics[width=\columnwidth]{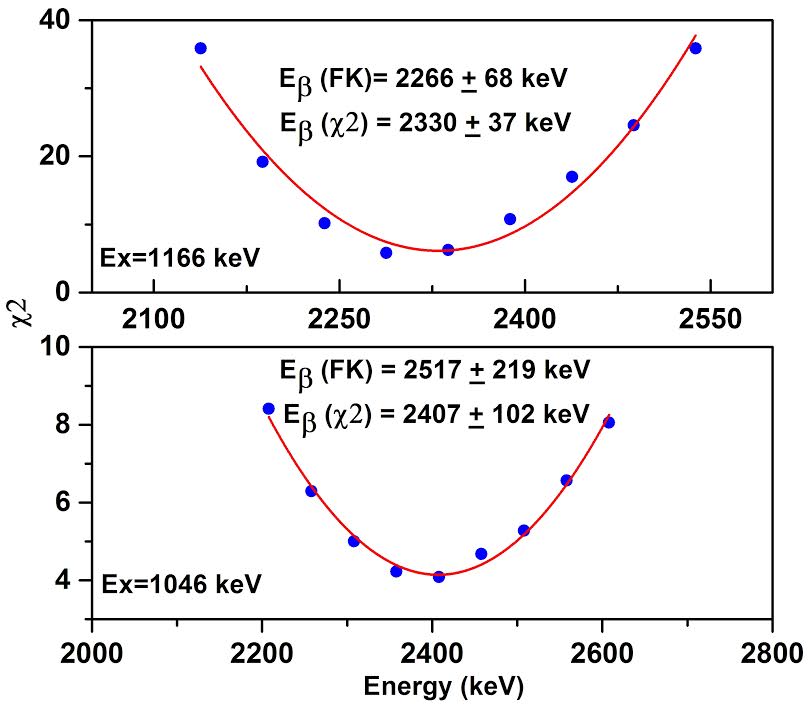}
  \caption{(Color Online) The representative $\chi ^2$ plot for two of the $\beta$ decay branches of $^{150g}$Pm. The $\beta$ decay end point energies obtained from $\chi ^2$ and FK analyses are indicated.}
  \label{fig-chi}
  \end{center}
  \end{figure}
\begin{figure}
  \begin{center}
  \hskip -1.0cm
  \includegraphics[width=\columnwidth]{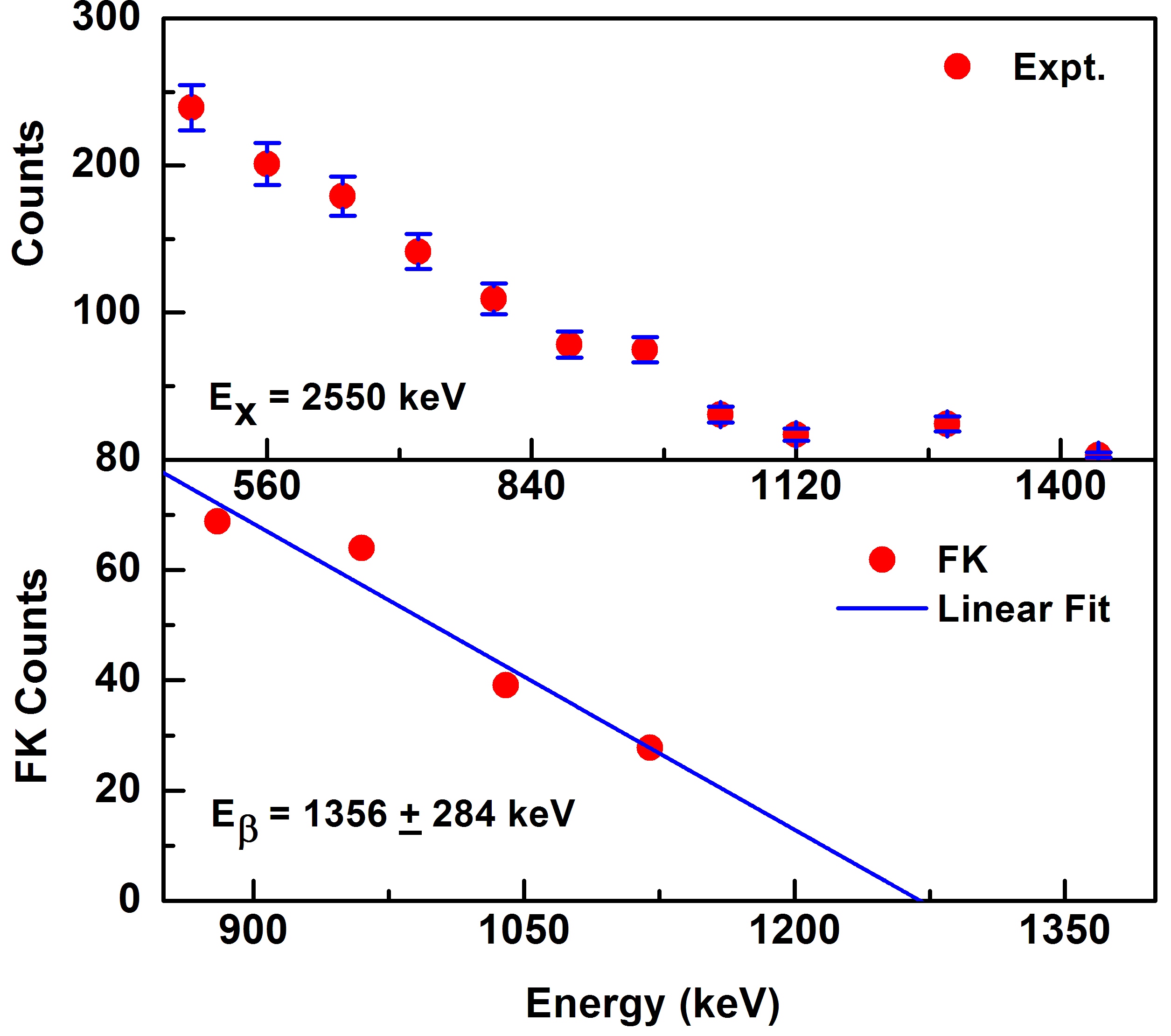}
  \caption{(Color Online) The $\beta$ decay spectrum and corresponding FK analysis for the beta decay of $^{150m}$Pm, the 2.2~h isomeric state.}
  \label{fig10a-beta}
  \end{center}
  \end{figure}
The endpoint energies have also been derived by performing a chi-square analysis following the methodology of Ref.~\cite{beta-tumpa} and the representative chi-square plots have been shown in Fig.~\ref{fig-chi}. The $\chi ^2$ values have been obtained by comparing the experimental data points with the simulation at different end point energies. The obtained values have been fitted with a parabolic function as shown in Fig.~\ref{fig-chi} and the minima of the obtained function has been considered as the value of the end-point energy. The end-point energies, thus, obtained have also been tabulated in table~\ref{Table-beta}. The error corresponds to the energy difference required for the change in $\chi ^2$ by 1.0.

In the present work, the measurement of $\beta$ decay endpoint energies have been performed for the first time in the decay of $^{150g}$Pm $\rightarrow$ $^{150}$Sm corresponding to several excited levels of the daughter. The $\beta -\gamma$ coincidence analyses have also been performed for obtaining the possible excitation energy of the isomeric level in $^{150}$Pm by using a higher binning of 80 keV/channel. The $\beta$ spectrum corresponding to the 1810 keV $\gamma$ gate have been shown in Fig.~\ref{fig10a-beta} which clearly shows that the obtained end point energy is higher than the value expected in case the transition had originated from the decay of $^{150g}$Pm. The FK analysis was performed and the $\beta$ end point energy obtained has been shown in the Fig.~\ref{fig10a-beta} and in table~\ref{Table-beta}. The excitation energy of the proposed isomeric state in $^{150}$Pm has been determined by comparing the $\beta$ decay Q value for $^{150g}$Pm, the excitation energy of the levels in $^{150}$Sm and the obtained end point energy. The values come out to be 453(284) keV from the above analysis. The $\beta$ spectrum corresponding to the other transitions originated from $^{150m}$Pm $\beta$ decay could not be determined in the present work, mainly due to low statistics.

\section{Systematics of long lived isomeric states around $^{150}$Pm and the possible structure of the proposed isomers in $^{150}$Pm}
\label{syst}

\begin{table*}
 \begin{center}
 \begin{small}
 \caption{Systematics of long lived isomers in odd-odd Pm nuclei neighboring to $^{150}$Pm}
\begin{tabular}{|cccc|ccc|ccc|}
 \multicolumn{4}{|c|}{Isomeric Level (odd-odd)}&\multicolumn{3}{c|}{proton~($\pi$)~level odd-Z}&\multicolumn{3}{c|}{neutron~($\nu$)~level odd-N}\\
\hline

 Nucleus&Ex. Energy&J$^{\pi}$&$\tau$~$\&$&Nucleus&Energy&J$^{\pi},\tau$&Nucleus&Energy&J$^{\pi},\tau$\\
 $\&$ Ref.&(keV)&Conf.&Decay&$\&$ Ref.&(keV)&Conf.&$\&$ Ref.&(keV)&Conf.\\

 \hline

&&&&&&&&&\\
 $^{140}$Pm&0.0&1$^+$&9.2s&$^{139}$Pm&0.0&(5/2$^+$)&$^{139}$Nd&0.0&3/2$^+$, 29.7m\\

 \cite{ensdf-a140}&&-&$\beta ^+ , \epsilon$&\cite{ensdf-a139}&&$\pi d_{\frac{5}{2}}$&\cite{ensdf-a139}&&-\\

 &(425)&8$^-$&5.95m&&188.7&(11/2$^-$), 180ms&&231.7&11/2$^-$,5.5h\\

 &&$\pi d_{\frac{5}{2}}\otimes \nu h_{\frac{11}{2}}$&$\beta ^+ , \epsilon$&&&$\pi \frac{3}{2} ^-$[541]&&&$\nu h_{\frac{11}{2}}$\\
&&&&&&&&&\\

 \hline

&&&&&&&&&\\
 $^{142}$Pm&0.0&1$^+$&40.5s&$^{141}$Pm&0.0&5/2$^+$&$^{141}$Nd&0.0&3/2$^+$, 2.49h\\
 \cite{ensdf-a142}&&-&$\beta ^+ , \epsilon$&\cite{ensdf-a141}&&$\pi d_{\frac{5}{2}}$&\cite{ensdf-a141}&&-\\

 &883.17&(8$^-$)&2.0ms&&196.87&7/2$^+$, 0.23ns&&193.68&1/2$^+$, 1.17ns\\

 &&-&IT&&&$\pi g_{\frac{7}{2}}$&&&-\\

 &&&&&628.62&11/2$^-$, 0.63$\mu$s&&756.51&11/2$^-$, 62s\\

 &&&&&&$\pi h_{\frac{11}{2}}$&&&$\nu h_{\frac{11}{2}}$\\

&&&&&&&&&\\

 \hline
 &&&&&&&&&\\
 $^{144}$Pm&0.0&5$^-$&363d&$^{143}$Pm &0.0&5/2$^+$, 265d&$^{143}$Nd&0.0&7/2$^-$\\
\cite{ensdf-a144}&&-&$\beta ^+ , \epsilon$&&&$\pi d_{\frac{5}{2}}$&&&$\nu f_{\frac{7}{2}}$\\
 &840.9&(9$^+$)&0.78$\mu$s&\cite{ensdf-a143}&272.04&7/2$^+$, 1.06ns&\cite{ensdf-a143}&742.05&3/2$^-$, 2.8ps\\
&&$\pi h_{\frac{11}{2}}\otimes \nu f_{\frac{7}{2}}$&IT&&&$\pi g_{\frac{7}{2}}$&&&\\
 &&&&&959.73&11/2$^-$, 24ns&&&\\
&&&&&&$\pi h_{\frac{11}{2}}$&&&\\
&&&&&&&&&\\
 \hline
&&&&&&&&&\\

 $^{148}$Pm&0.0&1$^-$&5.368d&$^{147}$Pm&0.0&7/2$^+$&$^{147}$Nd&0.0&5/2$^-$\\
\cite{ensdf-a148}&&-&$\beta ^-$&\cite{ensdf-a147}&&$\pi g_{\frac{7}{2}}$&\cite{ensdf-a147}&&-\\
 &137.9&5$^-$, 6$^-$&41.29d&&91.10&5/2$^+$, 2.5ns&&49.92&7/2$^-$, 1.0ns\\
 &&-&$\beta ^-$,IT&&&$\pi g_{\frac{7}{2}}$&&&$\nu h_{\frac{7}{2}}$\\
 &&&&&649.14&11/2$^-$, 27ns&&190.29&(9/2$^-$)\\
 &&&&&&$\pi h_{\frac{11}{2}}$&&&$\nu h_{\frac{9}{2}}$\\
&&&&&&&&&\\
 \hline
&&&&&&&&&\\

 $^{150}$Pm&0.0&1$^-$&2.698h&$^{149}$Pm&0.0&7/2$^+$&$^{149}$Nd&0.0&5/2$^-$\\
\cite{ensdf-a150}&&-&$\beta ^-$&&&$\pi \frac{7}{2}^+$[404]&&&$\nu \frac{5}{2} ^-$[523]\\
  $\&$&-&5$^-$&2.2h&\cite{ensdf-a149}&114.3&5/2$^+$, 2.53ns&\cite{ensdf-a149}&493.3&11/2$^-$\\
present&&-&$\beta ^-$,IT?&&&$\pi \frac{5}{2} ^+$[402]&&&$\nu \frac{11}{2}^-$[505]\\
 work&&&&&240.21&11/2$^-$, 35$\mu$s&&&\\
&&&&&&$\pi h_{\frac{11}{2}}$&&&\\
&&&&&&&&&\\
 \hline

 &&&&&&&&&\\
 $^{152}$Pm&0.0&1$^+$&4.12m&$^{151}$Pm&0.0&5/2$^+$, 28.4h&$^{151}$Nd&0.0&3/2$^+$, 12.44m\\
\cite{ensdf-a152}&&$\pi {\frac{5}{2}}^{-}[532]\otimes \nu {\frac{3}{2}}^{-}$[532]&$\beta ^-$&\cite{ensdf-a151}&&$\pi \frac{5}{2}^+$[413]&\cite{ensdf-a151}&&$\pi \frac{3}{2}^+$[651]+$\pi \frac{3}{2}^+$[402]\\
 &150&4$^-$&7.52m&&116.7&5/2$^-$, 89ps&&57.67&(3/2)$^-$\\
 &&$\pi {\frac{5}{2}}^{-}[532]\otimes \nu {\frac{3}{2}}^{-}$[532]&$\beta ^-$&&&$\pi \frac{5}{2}^-$[532]&&&$\nu \frac{3}{2}^-$[532]\\
 &150+x&8&13.8m&&255.6&3/2$^+$, 0.93ns&&189.054&(3/2)$^-$\\
&&-&$\beta ^-$,IT&&&$\pi \frac{3}{2}^+$[411]&&&$\nu \frac{3}{2}^-$[521]\\
 &&&&&&&&&\\

 \hline
&&&&&&&&&\\

 $^{154}$Pm&0&(3, 4)&2.68m&$^{153}$Pm&0.0&5/2$^-$&$^{153}$Nd&0.0&(3/2)$^-$\\
 \cite{ensdf-a154}&&$\pi {\frac{5}{2}}^{-}[532]\otimes \nu {\frac{3}{2}}^{-}$[521]&$\beta ^-$&&&$\pi \frac{5}{2}^-$[532]&&&$\pi \frac{3}{2}^{-}$[521]\\
&(210$\pm$70)&(0$^-$, 1$^-$)&1.73m&\cite{ensdf-a153}&32.194&3/2$^+$, 1.2ns&\cite{ensdf-a153}&191.7&(5/2$^-$)\\
 &&$\pi {\frac{5}{2}}^{-}[532]\otimes \nu {\frac{3}{2}}^{-}$[521]&$\beta ^-$&&&$\pi \frac{5}{2}^+$[413]&&&$\nu \frac{5}{2}^-$[523]\\
 &&&&&450.520&3/2$^+$&&&\\
 &&&&&&$\pi \frac{3}{2}^+$[411]&&&\\
&&&&&&&&&\\
 \hline
 &&&&&&&&&\\

 $^{156}$Pm&0.0&4($^+$)&26.70s&$^{155}$Pm &0.0&5/2$^-$&$^{155}$Nd&0.0&(3/2$^-$)\\
\cite{ensdf-a156,sood-pm156}&&$\pi {\frac{5}{2}}^{-}[532] + \nu {\frac{3}{2}}^{-}$[521]&$\beta ^-$&\cite{ensdf-a155}&180.565&5/2$^+$&\cite{ensdf-a153}&&\\
 &150.3&1$^+$&$<$5s&&&$\pi \frac{5}{2}^+$[413]&&&\\
&&$\pi {\frac{5}{2}}^{-}[532] - \nu {\frac{3}{2}}^{-}$[521]&IT,$\alpha$&&&&&&\\
&&&&&&&&&\\
 \hline

 \end{tabular}
  \label{tab.02}
 \end{small}
 \end{center}
 \end{table*}
 \begin{table*}
 \begin{center}
 \begin{small}
 \caption{Systematics of long lived isomers in N = 89 nuclei neighboring to $^{150}$Pm}
 \begin{tabular}{|cccc|ccc|ccc|}
 \hline

 \multicolumn{4}{|c|}{Isomeric Level (odd-odd)}&\multicolumn{3}{c|}{proton~($\pi$)~level odd-Z}&\multicolumn{3}{c|}{neutron~($\nu$)~level odd-N}\\
\hline

 Nucleus&Ex. Energy&J$^{\pi}$&$\tau$~$\&$&Nucleus&Energy&J$^{\pi},\tau$&Nucleus&Energy&J$^{\pi},\tau$\\
 $\&$ Ref.&(keV)&Conf.&Decay&$\&$ Ref.&(keV)&Conf.&$\&$ Ref.&(keV)&Conf.\\

 &&&&&&&&&\\
 \hline
 &&&&&&&&&\\

 $^{146}$La&0.0&(2$^-$)&6.1s&$^{145}$La&0.0&(5/2$^+$)&$^{145}$Ba&&\\

 \cite{ensdf-a146}&&-&$\beta ^-$,$\beta ^-$n&\cite{ensdf-a145}&-&&\cite{ensdf-a145}&&\\

 &(130)&(6$^-$)&9.8s&&572.4&(11/2$^-$)&&&\\

 &&-&$\beta ^-$,IT?&&-&&&&\\

 &&&&&&&&&\\
 \hline
 &&&&&&&&&\\
 $^{148}$Pr&0.0&1$^-$&2.29m&$^{147}$Pr&0.0&(5/2$^+$), 13.4m&$^{147}$Nd&0.0&5/2$^-$\\
\cite{ensdf-a148}&&$\pi {\frac{3}{2} ^+}[411]\otimes \nu {\frac{5}{2}}^-$[523]&$\beta ^-$&~\cite{ensdf-a147}&&$\pi \frac{5}{2}^+$[413]&~\cite{ensdf-a147}&&-\\
&76.80&4$^-$&2.01m&&&&&49.92&7/2$^-$, 1.0ns\\
&&$\pi {\frac{5}{2}}^{+}[413]\otimes \nu {\frac{3}{2}}^{-}$[532]&$\beta ^-$,IT&&&&&&$\nu f_{\frac{7}{2}}$\\
 &&&&&&&&190.29&(9/2$^-$)\\
&&&&&&&&&$\nu h_{\frac{9}{2}}$\\
 &&&&&&&&&\\
 \hline
 &&&&&&&&&\\

 $^{150}$Pm&0.0&1$^-$&2.698h&$^{149}$Pm&0.0&7/2$^+$&$^{149}$Nd&0.0&5/2$^-$\\
\cite{ensdf-a150}&&-&$\beta ^-$&\cite{ensdf-a149}&&$\pi \frac{7}{2}^+$[404]&\cite{ensdf-a149}&&$\nu \frac{5}{2} ^-$[523]\\
 $\&$&-&5$^-$&2.2h&&114.3&5/2$^+$, 2.53ns&&493.3&11/2$^-$\\
 present&&&$\beta ^-$,IT?&&&$\pi \frac{5}{2} ^+$[402]&&&$\nu \frac{11}{2}^-$[505]\\
 work&&&&&240.21&11/2$^-$, 35$\mu$s&&&\\
 &&&&&&$\pi h_{\frac{11}{2}}$&&&\\
 &&&&&&&&&\\
 \hline
 &&&&&&&&&\\

 $^{152}$Eu&0.0&3$^-$&13.517y&$^{151}$Eu&0.0&5/2$^+$&$^{151}$Sm&0.0&5/2$^-$\\
\cite{ensdf-a152}&&$\pi {\frac{5}{2}}^{+}[413]\otimes \nu {\frac{11}{2}}^{-}$[505]&$\epsilon$,$\beta ^+$,$\beta ^-$&\cite{ensdf-a151}&&$\pi d_{\frac{5}{2}}$&\cite{ensdf-a151}&&$\nu \frac{5}{2} ^-$[523]\\
 &45.59&0$^-$&9.31h&&21.541&7/2$^+$&&91.532&(9/2)$^+$\\
&&-&$\epsilon$,$\beta ^+$,$\beta ^-$&&&$\pi g_{\frac{7}{2}}$&&&$\nu i_{\frac{13}{2}}$\\
 &65.29&1$^-$&0.94$\mu$s&&&&&261.13&(11/2)$^-$\\
&&-&&&&&&&$\nu \frac{11}{2} ^-$[505]\\
 &65.29&1$^-$&0.94$\mu$s&&&&&&\\
 &&-&IT&&&&&&\\
 &147.86&8$^-$&96m&&&&&&\\
 &&$\pi {\frac{5}{2}}^{+}[413]\otimes \nu {\frac{11}{2}}^{-}$[505]&IT&&&&&&\\

 &&&&&&&&&\\
 \hline
 &&&&&&&&&\\

 $^{154}$Tb&0.0&0($^-$,$^+$)&21.5h&$^{153}$Tb&0.0&5/2$^+$&$^{153}$Gd&0.0&3/2$^-$\\
\cite{ensdf-a154}&&$\pi {\frac{3}{2}}^{+}[411]\otimes \nu {\frac{3}{2}}^{-}$[521]&$\epsilon$,$\beta ^+$,$\beta ^-$&\cite{ensdf-a153}&&$\pi \frac{5}{2} ^+$[402]&\cite{ensdf-a153}&&$\nu \frac{3}{2} ^-$[521]\\
 &&or&&&80.72&7/2$^+$, 0.49ns&&95.173&9/2$^+$, 3.5$\mu$s\\
&&$\pi {\frac{3}{2}}^{+}[411]\otimes \nu {\frac{3}{2}}^{+}$[651]&&&&$\pi \frac{7}{2} ^+$[404]&&&$\nu \frac{1}{2} ^+$[660]\\
 &$\le$25&3$^-$&9.4h&&147.57&(3/2$^+$), 0.84ns&&109.756&(5/2)$^-$,0.243ns\\
&&$\pi {\frac{3}{2}}^{+}[411]\otimes \nu {\frac{3}{2}}^{-}$[521]&IT,$\epsilon$,$\beta ^+$,$\beta ^-$&&&$\pi \frac{3}{2} ^+$[411]&&&$\nu \frac{5}{2} ^-$[523]\\
 &(200)&7$^-$&22.7h&&163.175&11/2$^-$, 186$\mu$s&&129.163&3/2$^-$, 2.52ns\\
&&$\pi {\frac{3}{2}}^{+}[411]\otimes \nu {\frac{11}{2}}^{-}$[505]&IT,$\epsilon$,$\beta ^+$&&&$\pi h_{\frac{11}{2}}$&&&$\nu \frac{3}{2} ^-$[532]\\
 &&&&&&&&171.188&(11/2$^-$), 76$\mu$s\\
&&&&&&&&&$\nu \frac{11}{2} ^-$[505]\\
&&&&&&&&&\\
 \hline
 &&&&&&&&&\\

 $^{156}$Ho&0.0&4$^-$&56m&$^{155}$Ho&0.0&5/2$^+$&$^{155}$Dy&0.0&3/2$^-$\\
  \cite{ensdf-a156}&&$\pi {\frac{5}{2}}^{+}[402]\otimes \nu {\frac{3}{2}}^{-}$[521]&$\epsilon$,$\beta ^+$&\cite{ensdf-a155}&&$\pi \frac{5}{2} ^+$[402]&\cite{ensdf-a155}&&$\nu \frac{3}{2} ^-$[521]\\
&52.37&1$^-$&9.5s&&110.16&7/2$^+$,$<$0.7ns&&132.195&9/2$^+$, 51ns\\
 &&$\pi {\frac{5}{2}}^{+}[402]\otimes \nu {\frac{3}{2}}^{-}$[521]&IT&&&$\pi \frac{7}{2} ^+$[404]&&&$\nu \frac{1}{2} ^+$[660]\\
 &(170)&9$^+$&7.6m&&141.97&11/2$^-$, 0.88ms&&136.319&5/2$^-$,$<$0.4ns\\
 &&$\pi {\frac{7}{2}}^{-}[523]\otimes \nu {\frac{11}{2}}^{-}$[505]&IT,$\epsilon$,$\beta ^+$&&&$\pi h_{\frac{11}{2}}$&&&$\nu \frac{5}{2} ^-$[523]\\
 &&&&&&&&202.413&3/2$^-$,$<$0.4ns\\
 &&&&&&&&&$\nu \frac{3}{2} ^-$[532]\\
 &&&&&&&&234.33&11/2$^-$, 6$\mu$s\\
 &&&&&&&&&$\nu \frac{11}{2} ^-$[505]\\
 &&&&&&&&&\\
 \hline
 \end{tabular}
  \label{tab.03}
 \end{small}
 \end{center}
 \end{table*}
Most of the nuclei around $^{150}$Pm show the presence of one or more long lived isomeric state(s). The systematics of these isomers around $^{150}$Pm have been shown in table~\ref{tab.02} and~\ref{tab.03}, where the isomeric levels having lifetime $\sim$$\mu$s or more have been considered. The systematics depict that these isomers share some common features. Most of these isomers have half lives greater than or close to that of the ground state and almost all of them except three undergo $\beta ^+$, $\beta ^-$ or EC decay. Also, in most of the cases, these isomers are negative parity levels and their spins are higher at least by an unit of 3$\hbar$ compared to the ground state spin. In table~\ref{tab.02} and~\ref{tab.03}, the known configurations to the isomeric states in the odd-odd nuclei around $^{150}$Pm are shown along with the available odd-proton and odd-neutron configurations taken from the respective odd-A neighbors.

 It is observed that the isomeric configurations in case of the odd-odd Pm nuclei are mainly of two types. For the Pm nuclei with N $\le$ 83, single particle configurations like $\pi d_{\frac{5}{2}}\otimes \nu h_{\frac{11}{2}}$ is proposed for the negative parity isomer in $^{140}$Pm and $\pi h_{\frac{11}{2}}\otimes \nu f_{\frac{7}{2}}$ configuration has been conjectured for the positive parity isomer in $^{144}$Pm. In case of more neutron rich Pm isotopes, these configurations are determined by the Nilsson configurations of $\pi \frac{5}{2}^{-}$[532] and $\nu \frac{3}{2}^{-}$[521], that correspond to the deformed $\pi 1h_{\frac{11}{2}}$ and $\nu 2f_{\frac{7}{2}}$ orbitals respectively, following the Quasi Particle Rotor Model~(QPRM) calculations~\cite{pqrm-sood}. In the systematics with N = 89, it is found that the configurations of these isomers in Z = 59 and Z = 63 are dominated by $\pi g_{\frac{7}{2}}\otimes \nu h_{\frac{11}{2}}$ configuration. In some of the cases with Z $\ge$ 65, however, the configurations involving $\nu 2f_{\frac{7}{2}}$ or both the  $\nu 1h_{\frac{11}{2}}$ and $\pi 1h_{\frac{11}{2}}$ are also seen to be existing.

 The above systematics clearly show that the isomeric configurations in the odd-odd nuclei can be conjectured by looking at the single particle configuration in the neighboring odd-Z and odd-N nuclei. Hence, the assignment of the configuration for the isomeric level in $^{150}$Pm can be attempted by taking the above understanding into consideration. The single particle configuration which is found to be active in the low lying states of odd-N nucleus neighboring to $^{150}$Pm, viz. $^{149}$Nd, is $\nu${$\frac{3}{2}^{-}$[523], the Nilsson configuration corresponding to deformed $\nu 2f_{\frac{7}{2}}$ orbital. In this nucleus, the state that corresponds to $\nu 1h_{\frac{11}{2}}$ is found at an excitation energy of 493 keV. In case of the odd-A Pm, viz., $^{149}$Pm and $^{151}$Pm, the single particle configurations close to the ground state are the Nilsson configurations corresponding to $\pi 1g_{\frac{7}{2}}$, $\pi 2d_{\frac{5}{2}}$ and $\pi 1h_{\frac{11}{2}}$. These are $\pi${$\frac{7}{2}^{+}$[404] in case of $^{149}$Pm ground state and $\pi${$\frac{5}{2}^{+}$[402] in case of 114.3 keV state in $^{149}$Pm. In $^{151}$Pm, the major configurations are known to be $\pi${$\frac{5}{2}^{+}$[413] for the $\frac{5}{2} ^+$ ground state, $\pi${$\frac{5}{2}^{-}$[532] for the $\frac{5}{2} ^-$, 117 keV state, $\pi${$\frac{3}{2}^{+}$[411] for the $\frac{3}{2} ^+$, 256 keV state and $\pi${$\frac{1}{2}^{+}$[420] for the $\frac{1}{2}^+$, 426 keV state. The first $\pi h_{\frac{11}{2}}$ level in $^{149}$Pm lies at 240 keV and is also an isomeric state. In case of $^{151}$Pm this proton configuration is observed at much lower excitation of 117 keV.

 Considering the available single particle orbitals around $^{150}$Pm, it is understood that the involvement of $\pi h_{\frac{11}{2}}$ particle or a $\nu h_{\frac{11}{2}}$ hole in the configurations of the proposed isomeric state will make the excitation energy of the isomer very high. Hence, it may be conjectured that the low lying isomers with high spin may be one of the multiplets generated by involving the $\pi g_{\frac{7}{2}}$ and the  $\nu f_{\frac{7}{2}}$ orbitals which give rise to the negative parity levels. There exist no theoretical model calculation to understand the excited energy levels of $^{150}$Pm. In the present work, a shell model calculation has been performed by using OXBASH code~\cite{BAB94} to examine the negative parity isomers in $^{150}$Pm, as described in the following section~\ref{shell}, with a restricted particle configuration in order to reproduce the low lying negative parity levels in the nucleus.

\section{Shell Model Calculation}
\label{shell}

\begin{figure}
  \begin{center}
  \hskip -1.0cm
  \includegraphics[width=\columnwidth]{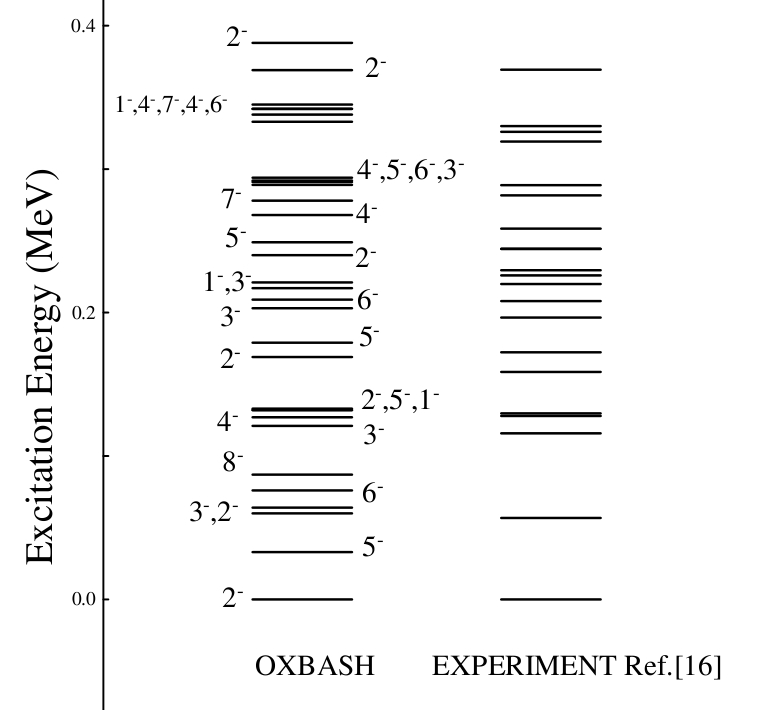}
  \caption{The level energies for the excited levels of $^{150}$Pm are compared with shell model calculation. The J$^{\pi}$ values corresponding to the each levels have been obtained from calculation.}
  \label{shmod}
  \end{center}
  \end{figure}
In order to find the structure of the low lying states in $^{150}$Pm, a Large Basis Shell Model (LBSM) calculation was performed using the code OXBASH \cite{BAB94}. The calculation considered
$^{132}$Sn as core and eleven protons were distributed over the model space
comprising of $\pi$(1$g_{\frac{7}{2}}$, 2$d_{\frac{5}{2}}$, 2$d_{\frac{3}{2}}$, 3$s_{\frac{1}{2}}$, 1h$_{\frac{11}{2}}$) single particle orbitals along with seven neutrons over the $\nu$(1$h_{\frac{9}{2}}$, 2$f_{\frac{7}{2}}$, 2$f_{\frac{5}{2}}$, 3$p_{\frac{3}{2}}$, 3$p_{\frac{1}{2}}$, 1i$_{\frac{13}{2}}$)
single particle orbitals. The calculations were carried out using
proton-neutron formalism in full valence space applying particle restriction as shown in Table~\ref{conf}. This choice was based on both the facts that increasing the number of combinations for the single particle orbitals were not possible due to increasing dimension of the matrix and also the aim of the present work was to reproduce the negative parity levels in comparison to the negative parity ground state. The two-body matrix elements were obtained from the well-known $\it{cwg}$ interaction~\cite{cwg} supplied with the code OXBASH. The calculated excitation energies of the negative parity states up to 300 keV excitation have been shown in Fig.~\ref{shmod}, along with the experimental levels obtained via (p,n$\gamma$) reaction in the work of Bucurescu {\it et al}~\cite{pm150-prc}. A level by level comparison was not possible as no experimental spin parity assignment for these levels exists in literature. The comparison clearly indicates that the ground state of $^{150}$Pm is likely to be 2$^-$ which has been reported as 1$^-$ in ENSDF~\cite{ensdf-a150} and suggested as 2$^-$ in the work of D. Bucurescu et al~\cite{pm150-prc}. From the comparison, it is also observed that there is one 5$^-$ state predicted at 33 keV excitation which lies between the 2$^-$ ground state and the $3^- _1$ excited state. The major configurations of all the calculated levels were studied and it is observed that although the proton configuration for both the ground state and the 5$^-$ excited state is similar, the ground state wave function is dominated by the $\nu$2$f_{\frac{7}{2}}$ configuration whereas contributions from both the $\nu$3$p_{\frac{3}{2}}$ and $\nu$2$f_{\frac{7}{2}}$ orbitals are found to exist as the major configurations of the 5$^-$ state. However, the contribution of the configuration involving $\nu$1$h_{\frac{9}{2}}$ in both the ground state and the 5$^-$ state is calculated to be less than 1$\%$.

In the present work, the transition probability for the above mentioned 5$^-$ level to the 2$^-$ ground state was calculated by considering the resultant 33 keV $\gamma$ transition as either of E4 or M3 in nature. The said transition probabilities come out to be 51.4 $\mu _N^2fm^3$ for the M3 transition and 8545 $e^2fm^8$ for the E4 transition respectively. These values correspond to 1.3$\times 10^4$ years (M3) and 2.5$\times 10^5$ years (E4) of lifetimes respectively for $\gamma$ decay from the predicted 5$^-$ level. These values corresponding to the 150 keV energy for the isomer yield lifetime of 5.7 days (M3) and 31 days (E4) respecively. However, while considering an energy value of 454 keV for the isomer the above lifetimes go down to 3.5 m (M3) and 2.1 m (E4) respectively. The Weisskopf estimates for these respective M3 and E4 transition rates were also calculated and comes out about 1315 $\mu _N^2fm^3$ (M3) and 39500 $e^2fm^8$ (E4) respectively. These, in turn, corresponds to the $\gamma$ decay half lives as 5.0$\times 10^2$ years (M3) and 5.4$\times 10^4$ years (E4) respectively. Hence, both from the half life values obtained from shell model as well as from single particle estimates, it can be concluded that the lowest 5$^-$ level in $^{150}$Pm will have almost no $\gamma$ decay to the 2$^-$ ground state of $^{150}$Pm. Thus, the above calculations indicate the presence of a 5$^-$, 33 keV isomer in $^{150}$Pm which undergoes $\beta ^-$ decay to the excited states of $^{150}$Sm nucleus, as shown in Fig.~\ref{fig9-isomer2.2}.

\begin{table*}
\caption{The particle restrictions applied for the shell model calculation of $^{150}$Sm levels. The possible configurations those may be responsible for the generation of +ve and -ve parity levels are indicated.}
\begin{center}
\begin{tabular}{|cccc||cc|cc|}
\hline
\hline
\multicolumn{4}{|c||}{Particle Restriction}&\multicolumn{4}{c|}{Possible configuration}\\
\hline
&&&&&&&\\
&Orbitals&Minimum&Maximum&\multicolumn{2}{c|}{Proton($\pi$)}&\multicolumn{2}{c|}{Neutron($\nu$)}\\
&&&&&&&\\
&&&&\multicolumn{1}{c|}{$(g_{\frac{7}{2}},d_{\frac{5}{2}},d_{\frac{3}{2}},s_{\frac{1}{2}})$}&$(h_{\frac{11}{2}})$&\multicolumn{1}{c|}{$(f_{\frac{7}{2}},h_{\frac{9}{2}},f_{\frac{5}{2}}, p_{\frac{3}{2}},p_{\frac{1}{2}})$}&$(i_{\frac{13}{2}})$\\
&&&&\multicolumn{1}{c|}{}&&\multicolumn{1}{c|}{}&\\
\hline
\multicolumn{8}{|c|}{}\\
\multicolumn{8}{|c|}{Nucleus:$^{150}$Pm}\\
\multicolumn{8}{|c|}{}\\
\hline
&&&&\multicolumn{1}{c|}{}&&\multicolumn{1}{c|}{}&\\
&&&&\multicolumn{1}{c|}{-ve parity states~:}&&\multicolumn{1}{c|}{}&\\
Proton ($\pi$)&2$d_{\frac{5}{2}}$&2&3&\multicolumn{1}{c|}{}&&\multicolumn{1}{c|}{}&\\
&2$d_{\frac{3}{2}}$&0&1&\multicolumn{1}{c|}{1}&\multicolumn{1}{c|}{0}&\multicolumn{1}{c|}{1}&0\\
&3$s_{\frac{1}{2}}$&0&1&\multicolumn{1}{c|}{0}&1&\multicolumn{1}{c|}{0}&1\\
&1$h_{\frac{11}{2}}$&0&1&\multicolumn{1}{c|}{}&\multicolumn{1}{c|}{}&\multicolumn{1}{c|}{}&\multicolumn{1}{c|}{}\\
&1$h_{\frac{11}{2}}$&0&1&\multicolumn{1}{c|}{+ve parity states~:}&&\multicolumn{1}{c|}{}&\\
&&&&\multicolumn{1}{c|}{}&&\multicolumn{1}{c|}{}&\\
Neutron($\nu$)&2$f_{\frac{7}{2}}$&6&7&\multicolumn{1}{c|}{1}&\multicolumn{1}{c|}{0}&\multicolumn{1}{c|}{0}&1\\
&1$h_{\frac{9}{2}}$&0&1&\multicolumn{1}{c|}{0}&\multicolumn{1}{c|}{1}&\multicolumn{1}{c|}{1}&0\\
&2$f_{\frac{5}{2}}$&0&1&\multicolumn{1}{c|}{}&\multicolumn{1}{c|}{}&\multicolumn{1}{c|}{}&\multicolumn{1}{c|}{}\\
&3$p_{\frac{3}{2}}$&0&1&\multicolumn{1}{c|}{}&\multicolumn{1}{c|}{}&\multicolumn{1}{c|}{}&\multicolumn{1}{c|}{}\\
&3$p_{\frac{1}{2}}$&0&1&\multicolumn{1}{c|}{}&\multicolumn{1}{c|}{}&\multicolumn{1}{c|}{}&\multicolumn{1}{c|}{}\\
&1$i_{\frac{13}{2}}$&0&1&\multicolumn{1}{c|}{}&\multicolumn{1}{c|}{}&\multicolumn{1}{c|}{}&\multicolumn{1}{c|}{}\\
&&&&\multicolumn{1}{c|}{}&\multicolumn{1}{c|}{}&\multicolumn{1}{c|}{}&\multicolumn{1}{c|}{}\\
\hline
\multicolumn{8}{|c|}{}\\
\multicolumn{8}{|c|}{Nucleus:$^{150}$Sm}\\
\multicolumn{8}{|c|}{}\\
\hline
&&&&\multicolumn{1}{c|}{}&&\multicolumn{1}{c|}{}&\\
&&&&\multicolumn{1}{c|}{-ve parity states~:}&&\multicolumn{1}{c|}{}&\\
Proton ($\pi$)&1$g_{\frac{7}{2}}$&7&8&\multicolumn{1}{c|}{1}&\multicolumn{1}{c|}{1}&\multicolumn{1}{c|}{2}&0\\
&2$d_{\frac{5}{2}}$&3&4&\multicolumn{1}{c|}{2}&\multicolumn{1}{c|}{0}&\multicolumn{1}{c|}{1}&1\\
&2$d_{\frac{3}{2}}$&0&1&\multicolumn{1}{c|}{1}&\multicolumn{1}{c|}{1}&\multicolumn{1}{c|}{0}&2\\
&3$s_{\frac{1}{2}}$&0&1&\multicolumn{1}{c|}{0}&\multicolumn{1}{c|}{2}&\multicolumn{1}{c|}{1}&1\\
&1$h_{\frac{11}{2}}$&0&1&\multicolumn{1}{c|}{}&&\multicolumn{1}{c|}{}&\\
&&&&\multicolumn{1}{c|}{+ve parity states~:}&&\multicolumn{1}{c|}{}&\\
Neutron ($\nu$)&1$h_{\frac{9}{2}}$&0&1&\multicolumn{1}{c|}{}&&\multicolumn{1}{c|}{}&\\
&2$f_{\frac{7}{2}}$&5&6&\multicolumn{1}{c|}{2}&\multicolumn{1}{c|}{0}&\multicolumn{1}{c|}{2}&0\\
&2$f_{\frac{5}{2}}$&0&1&\multicolumn{1}{c|}{0}&\multicolumn{1}{c|}{2}&\multicolumn{1}{c|}{0}&2\\
&3$p_{\frac{3}{2}}$&0&1&\multicolumn{1}{c|}{1}&\multicolumn{1}{c|}{1}&\multicolumn{1}{c|}{1}&1\\
&3$p_{\frac{1}{2}}$&0&1&\multicolumn{1}{c|}{}&&\multicolumn{1}{c|}{}&\\
&1$i_{\frac{13}{2}}$&0&1&\multicolumn{1}{c|}{}&&\multicolumn{1}{c|}{}&\\
&&&&\multicolumn{1}{c|}{}&&\multicolumn{1}{c|}{}&\\
\hline
\end{tabular}
\end{center}
\label{conf}
\end{table*}

As pointed out in Section~\ref{i-logft}, the 5$^-$ isomeric level could have various types of $\beta$ decay modes, viz., {\it the allowed Gammow-Teller (GT), first Forbidden (FF), FF unique and second forbidden transition}. We attempted, using OXBASH, to calculate the {\it logft} values for the {\it allowed GT} transitions (characterised by $\Delta l$ = 0, $\Delta J$ = 0,1 (no 0 $\rightarrow$ 0), $\Delta \pi$ = 0) from the 2$^-$ ground state and 5$^-$ isomeric state of $^{150}$Pm. For this calculation, first the energy levels of $^{150}$Sm have been calculated by using the same formalism, core nucleus and the interaction matrix elements as used for $^{150}$Pm, discussed above. The particle restriction was required to truncate the model space so that the calculation could be performed with permitted dimension of the matrix. The available orbitals, the applied particle restriction and the possible configurations for the positive and negative parity levels in $^{150}$Sm have been given in table~\ref{conf}.
Although the said calculation could reproduce some of the low lying positive parity levels, the excitation energies of the negative parity levels in $^{150}$Sm could not be reproduced well. This may be due to the truncation of the model space which forced in neglecting the configurations that associate more than one particle in many of the orbitals, importantly in the unique parity $\pi 1h_{\frac{11}{2}}$ and $\nu 1i_{\frac{13}{2}}$ orbitals, as understood from the listed possible configurations in table~\ref{conf}. With this limitation in the calculation, the configuration of the 3$^-$, 4$^-$ and 6$^-$ levels in $^{150}$Sm were found to have been generated from one proton particle in $\pi 1h_{\frac{11}{2}}$. As per our particle restriction, this configuration is the first negative parity configuration mentioned in table~\ref{conf}. Now, any {\it GT} type $\beta^{-}$ decay from the ground state or the 5$^-$ isomeric state in $^{150}$Pm to these negative parity levels in $^{150}$Sm is possible only by the conversion of one $\nu 1h_{\frac{9}{2}}$ neutron to a $\pi 1h_{\frac{11}{2}}$ proton. For brevity, authors do not want to give details of these calculations, except to mention that the allowed GT decays from the 5$^-$ isomeric state to the first ten 4$^-$,5$^-$ and 6$^-$ states in $^{150}$Sm have {\it logft} values in the range of 7 - 11 and for the decay of the 2$^-$ ground state to the first and second 3$^-$ state of $^{150}$Sm have {\it logft} values 9.1 and 8.2 respectively. The measured {\it logft} ($2^-_{g.s} \rightarrow 3^-_2$) is 9.65. Hence, this effort at least indicates the possibility of the $\beta ^-$ decay of the 5$^-$ isomeric level and confirms its presence in the level scheme of $^{150}$Pm.

\section{Summary}

The decay spectroscopy of $^{150}$Pm has been performed by populating the nucleus with $^{150}$Nd(p,n$\gamma$)$^{150}$Pm reaction at E$_p$ = 8.0 MeV. The obseved $\gamma$ rays were counted both in singles and coincidence mode with the VENUS array having six Compton suppressed Clover HPGe detectors. Following the decay curves of the observed transitions and their $\gamma -\gamma$ coincidence relationship could confirm the presence of a long lived $\beta$-decaying isomer in $^{150}$Pm having half life 2.2(1)~h. The {\it logft} analyses were performed using the intensities of different $\gamma$ transitions which suggest modified J$^{\pi}$ assignment for the levels in $^{150}$Sm that are fed by the newly identified isomeric level. The $\beta$ decay end point energies, corresponding to several decay branches of $^{150}$Pm ground state, have been measured for the first time by using the $\beta -\gamma$ coincidence technique with an array of two thin window LEPS and four Clover HPGe detectors of the VENUS array. The end point energy measurement for the isomeric $\beta$ decay indicates the level energy of the isomer as 453(284) keV. The systematics of the configurations associated with the long lived isomers, neighboring to $^{150}$Pm, have been studied. Shell model calculation has been performed by using OXBASH code that clearly indicates a 5$^-$ isomeric level at 33 keV excitation of $^{150}$Pm. The configuration of the isomer is suggested to be a mixture of $\pi (1g_{\frac{7}{2}},2d_{\frac{5}{2}})\otimes \nu 3p_{\frac{3}{2}}$ and $\pi (1g_{\frac{7}{2}},2d_{\frac{5}{2}})\otimes \nu 2f_{\frac{7}{2}}$ configurations. The transition probabilities calculated for this state to the 2$^-$ ground state, having a $\pi (1g_{\frac{7}{2}},2d_{\frac{5}{2}})\otimes \nu 2f_{\frac{7}{2}}$ configuration, suggest that the electromagnetic decay half life of the isomer is greater than 10$^4$ y and undergoes $\beta$ decay. The contribution of the single particle configuration involving $\nu$1$h_{\frac{9}{2}}$ orbital is conjectured to be present both in the ground state and the 5$^-$ isomeric state which makes the {\it allowed GT} type $\beta$ decay possible from these two levels in $^{150}$Pm. The confirmation of the existence of 5$^-$ isomeric level in $^{150}$Pm level scheme and the possibility of its $\beta$ decay were also obtained from the calculation of {\it logft} values using OXBASH. The absence of experimental data for the $^{150}$Pm nucleus warrants a further in-beam $\gamma$ spectroscopic measurement which has been planned as a future experimental outlook to this work. A detailed LBSM calculation for this valence space is really challenging.

\section{Acknowledgement}
\label{ack}

The effort of the staffs and members of the K=130 cyclotron operation group at VECC, Kolkata, is gratefully acknowledged for providing high quality stable proton beam. A. Saha is grateful for his UGC Fellowship (Ref.~No:17-06/2012(i)EU-V) for carrying out his experiments at VECC, Kolkata. S. S. Alam would like to acknowledge the support from BRNS fellowship (Sanction No. 2013/38/02-BRNS/1927 for PRF, BRNS, dated 16 October 2013 ). The efforts of Mr. R. K. Chatterjee, RCD, VECC is acknowledged for target preparation. A. Chowdhury and Shaikh Imran of Physics lab, VECC are acknowledged for their effort to maintain the detectors during experiment.

\end{document}